\documentclass[aps,prd,reprint,amsmath,amssymb,nobibnotes,showpacs,a4paper]{revtex4-1}

\printfigures
\usepackage{amsfonts}
\usepackage{bm}
\usepackage[pdftex]{graphicx}
\usepackage{dcolumn}
\usepackage{latexsym}
\usepackage{longtable}
\usepackage{color}
\begin{document}
\title{Kinematic state of an interacting cosmology modeled with Chebyshev polynomials}
\author{Freddy Cueva Solano}
\affiliation{Instituto de F\'{\i}sica y Matem\'aticas, Universidad Michoacana de San Nicol\'as de Hidalgo\\
Edificio C-3, Ciudad Universitaria, CP. 58040, Morelia, Michoac\'an, M\'exico.}
\email{freddy@ifm.umich.mx,\;\;freddycuevasolano$2009$@gmail.com}
\date{\today}
\begin{abstract}
In a spatially flat universe and for an interacting cosmology, we have reconstructed the interaction term, $Q$, between a cold dark 
matter (DM) fluid and a dark energy (DE) fluid, as well as a time-varying equation of state (EoS) parameter $\mathrm{\omega_{DE}}$, and have explored 
their cosmological impacts on the amplitudes of the first six cosmographic parameters, which allow us to extract information about the kinematic state of the 
universe today. Here, both $Q$ and $\mathrm{\omega_{DE}}$ have been modeled in terms of the Chebyshev polynomials. Then, via a Markov-Chain Monte Carlo 
(MCMC) method, we have constrained the model parameter space by using a combined analysis of geometric data. Our results show that the evolution curves 
of the cosmographic parameters deviate strongly from those predicted in the standard model when are compared, namely, they are much more sensitive to 
$Q$ and $\mathrm{\omega_{DE}}$ during their cosmic evolution. In this context, we have also found that different DE scenarios could be compared and 
distinguished among them, by using the present values of the highest order cosmographic parameters.
\end{abstract}
\maketitle 
\section{Introduction}  
Nowadays, a huge number of independent observational evidences 
\cite{Conley2011,Jonsson2010,Betoule2014,Planck2015,Hinshaw2013,Beutler2011,Ross2015,Percival2010,Blake2011,Kazin2010,Anderson2014a,Padmanabhan2012,Chuang2013a,Chuang2013b,Debulac2015,FontRibera2014,Eisenstein1998,Eisenstein2005,Hemantha2014,Bond-Tegmark1997,Hu-Sugiyama1996,Neveu2016,Zhang2014,Simon2005,Moresco2012,Moresco2016,Gastanaga2009,Oka2014,Blake2012,Stern2010,Moresco2015,Busca2013} 
reveal that the universe is undergoing an accelerated expansion during the late times cosmic. This observed phenomenon is a transcendental issue today in 
Cosmology and its understanding from physical arguments is still unclear. In this sense, two kinds of explanations can describe that phenomenon, but 
they are different in nature. The first one requires the existence of an exotic form of energy with negative pressure usually called DE \cite{DES2006}. 
This energy has been interpreted in various forms and widely studied in \cite{OptionsDE}. The second one is the large-distance 
modification of gravity, which leads to the cosmic acceleration today \cite{modifiedDE}. Due to the degeneracy between the space of parameter and the 
cosmic expansion, it is difficult to decide which above explanation is correct.\\
In the literature, an alternative way have been proposed without the use of cosmological variables coming from dynamical descriptions and under the 
assumption that the Friedmann-Robertson-Walker (FRW) metric is still valid. This approach is denominated Cosmography or Cosmokinetics. Hence, in general, 
via the Taylor expansion of the scale factor $a(t)$, in terms of the cosmic time $t$, we truncate the series at the sixth order and the 
dimensionless coefficients defined today $\mathrm{q_{0}}$, $\mathrm{j_{0}}$, $\mathrm{s_{0}}$, $\mathrm{l_{0}}$ and $\mathrm{m_{0}}$ are respectively 
denominated as deceleration, jerk, snap, lerk and merk, and called cosmographic parameters, which describe the kinematic state of the universe 
\cite{Weinberg1972,Visser2005,Xu-Li-Lu2009,Wang-Dai-Qi2009,Vitagliano-Viel2010,Xu2011,Gruber2012,Sendra2013,Gruber2014,Movahed2017,Pan2018,Ming2017,Rodrigues2018,Ruchika2018} 
and can be measured by cosmic observations today.\\
On the other one, an interacting DE model (IDE) with two different cases is discussed here. Due to the lacking of an underlying theory for construct a 
general term of interaction, $Q$, between the dark sectors, different ansatzes have been widely discussed in 
\cite{Interacting,Pavons,Wangs,Cueva-Nucamendi2012,valiviita2008,Clemson2012}. So, It has been shown in DE scenarios that $Q$ can affect the background 
the expansion history of the universe and could very possibly introduce new features on the evolution curves of the cosmographic parameters.
In this letter, we have attempted phenomenological descriptions for $Q$ and $\mathrm{\omega_{DE}}$, by expanding them in terms of the 
Chebyshev polynomials $T_{n}$, defined in the interval $[-1,1]$ and with a divergence-free $\mathrm{\omega_{DE}}$ at 
$z\rightarrow -1$ \cite{Chevallier-Linder,Li-Ma}. However, that polynomial base was particularly chosen due to its rapid convergence and better 
stability than others, by giving minimal errors \cite{Simon2005,Martinez2008}. Besides, $Q$ could also be proportional to the DM energy density 
$\mathrm{{\rho}_{DM}}$ and to the Hubble parameter $\mathrm{\bf H}$. Here, $Q$ will be restricted from the criteria exhibit in \cite{Campo-Herrera2015}.\\
The focus of this paper is to investigate the effects of $Q$ and $\mathrm{\omega_{DE}}$ on the evolution of the first six 
cosmographic parameters and compare them with the results of non-interacting models.\\
To constrain the parameter spaces of our models, break the degeneracy of their parameters and put tighter constraints on them, we use an analysis combined of 
Joint Light Curve Analy-sis (JLA) type Ia Supernovae (SNe Ia) data \cite{Conley2011,Jonsson2010,Betoule2014}, including with Baryon Acoustic Oscillation (BAO) data 
\cite{Hinshaw2013,Beutler2011,Ross2015,Percival2010,Blake2011,Kazin2010,Anderson2014a,Padmanabhan2012,Chuang2013a,Chuang2013b,Debulac2015,FontRibera2014}, 
together the Planck distance priors of the Cosmic Microwave Background (CMB) data, 
\cite{Planck2015,Bond-Tegmark1997,Hu-Sugiyama1996,Neveu2016} and the Hubble parameter ($\bf H$) data obtained from galaxy 
surveys \cite{Zhang2014,Simon2005,Moresco2012,Moresco2016,Gastanaga2009,Oka2014,Blake2012,Stern2010,Moresco2015,Busca2013}.\\
The main result that we have found here, is that, the amplitudes of the cosmographic parameters in the IDE model deviate significantly of those inferred in the 
non-interacting models. It could be used to establish differences among our models.\\
The paper is organized as follows. In Sec. 2, we have described the phenomenological model considered here. In Sec. 3, we have presented the cosmographic parameters. 
In Sec. 4 provides a description of the constraint method and observational data. We discuss the results obtained in Sec. 5. Finally, we have summarized the 
conclusions In Sec. 6.
\section{Interacting dark energy (IDE) model}\label{Background}
We assume a spatially flat FRW universe, composed with four perfect fluids-like, radiation (subscript r), baryonic matter (subscript b), 
DM and DE, respectively. Moreover, we postulate the existence of a non-gravitational coupling in the background between DM and DE (so-called dark sector) 
and two decoupled sectors related to the b and r components, respectively. We also consider that these fluids have EoS parameters 
${\mathrm{P}}_{A}=\mathrm{\omega}_{A}\mathrm{{\rho}}_{A}$, $A=b,r,DM,DE$, where $\mathrm{P}_{A}$ and $\mathrm{{{\rho}}}_{A}$ are the 
corresponding pressures and the energy densities. Here, we choose $\mathrm{\omega}_{DM}=\mathrm{\omega}_{b}=0$, 
$\mathrm{\omega}_{r}=1/3$ and $\mathrm{\omega}_{DE}$ is a time-varying function. Therefore, the balance equations of our fluids are respectively, 
\begin{eqnarray}
\label{EB}\frac{\mathrm{d}{\rho}_{b}}{\mathrm{d}z}-3{\mathrm{\bf H}}{{\rho}}_{b}&=&0,\\
\label{Er}\frac{\mathrm{d}{\rho}_{r}}{\mathrm{d}z}-4{\mathrm{\bf H}}{{\rho}}_{r}&=&0,\\
\label{EDM}\frac{\mathrm{d}{\rho}_{DM}}{\mathrm{d}z}-\frac{3{{\rho}}_{DM}}{(1+z)}&=&-\frac{Q}{\mathrm{\bf H}(1+z)},\\
\label{EDE}\frac{\mathrm{d}{\rho}_{DE}}{\mathrm{d}z}-\frac{3(1+\omega_{DE}){{\rho}}_{DE}}{(1+z)}&=&+\frac{Q}{\mathrm{\bf H}(1+z)},
\end{eqnarray}
where the differentiation has been done with respect to the redshift, $z$, $\mathrm{\bf H}$ denotes the Hubble expansion rate and the quantity $Q$ expresses 
the interaction between the dark sectors. For simplicity, it is convenient to define the fractional energy densities 
$\mathrm{\Omega}_{A}\equiv\frac{\mathrm{\rho}_{A}}{\rho_{c}}$ and $\mathrm{{\Omega}_{A,0}}\equiv\frac{{\rho}_{A,0}}{\rho_{c,0}}$, where the 
critical density $\rho_{c}\equiv 3{\mathrm{\bf H}}^2/8\pi G$ and the critical density today $\rho_{c,0}\equiv 3\mathrm{H_{0}}^{2}/8\pi G$ being 
$\mathrm{H_{0}}=100h\,Kms^{-1}Mpc^{-1}$ the current value of $\mathrm{\bf H}$. Likewise, we have taken the relation $\sum_{A}{{\mathrm{\Omega}}_{A,0}}=1$. Here, 
the subscript ``0'' indicates the present value of the quantity.\\ 
In this work, we consider the spatially flat FRW metric with line element 
\begin{equation}\label{metricbackground} 
{\rm ds}^{2}=-{\rm d}{t}^{2}+{\mathrm{\bf a}}^{2}(t)\delta_{ij}d{x}^{i}d{x}^{j},
\end{equation}
where $t$ represents the cosmic time and ``$\mathrm{\bf a}$'' represents the scale factor of the metric and it is defined in terms of the 
redshift $\mathrm{z}$ as $a=(1+z)^{-1}$, from which one can find the relation of $\mathrm{\bf H}$ and the cosmic time 
$dt/dz=-1/(1+z)\mathrm{\bf H}$.\\
Then, we analyze the ratio between the energy densities of DM and DE, defined as $\mathrm{R}\equiv \mathrm{{\rho}_{DM}}/\mathrm{{\rho}_{DE}}$. 
From Eqs. (\ref{EDM}) and (\ref{EDE}), we obtain \cite{Campo-Herrera2015,Ratios}
\begin{equation}\label{ratio}
\frac{\mathrm{dR}}{\mathrm{d}z}=\frac{-\mathrm{R}}{(1+z)}\left(3\omega_{DE}+\frac{(1+\mathrm{R})Q}{\mathrm{\bf H}\rho_{DM}}\right).
\end{equation}
This Eq. leads to 
\begin{equation}\label{FormQ}
Q=-\left(3\omega_{DE}+\frac{\mathrm{dR}}{\mathrm{d}z}\frac{(1+z)}{\mathrm{R}}\right)\frac{\mathrm{\bf H}{{\rho}}_{DM}}{1+\mathrm{R}}.
\end{equation}
Due to the fact that the origin and nature of the dark fluids are unknown, it is not possible to derive $Q$ from fundamental principles. 
However, we have the freedom of choosing any possible form of $Q$ that satisfies Eqs. (\ref{EDM}) and (\ref{EDE}) simultaneously. Hence, we 
propose a phenomenological description for $Q$ as a linear combination of $\mathrm{{{\rho}}_{DM}}$, ${\mathrm{\bf H}}$ and a time-varying 
function ${{\rm I}}_{\rm Q}$,
\begin{equation}\label{Interaction}
{Q}\equiv \mathrm{\bf H}\mathrm{{\rho}_{DM}}{{\rm I}}_{\rm Q},\qquad {{\rm I}}_{\rm Q}\equiv \sum_{n=0}^{}{{\lambda}}_{n}T_{n},
\end{equation}
where ${{\rm I}}_{\rm Q}$ is defined in terms of Chebyshev polynomials and $\mathrm{{\lambda}_{n}}$ are constant and small $|\mathrm{{\lambda}_{n}}|\ll1$ 
dimensionless parameters. This polynomial base was chosen because it converges rapidly, is more stable than others and behaves well in any polynomial 
expansion, giving minimal errors \cite{Cueva-Nucamendi2012}. The first three Chebyshev polynomials are
\begin{equation}\label{Chebyshev1} 
T_{0}(z)=1\;,\hspace{0.3cm}T_{1}(z)=z\;,\hspace{0.3cm} T_{2}(z)=(2z^{2}-1).
\end{equation}
From Eqs. (\ref{Interaction}) and (\ref{Chebyshev1}) an asymptotic value for ${{\rm I}}_{\rm Q}$ can be found:
${{\rm I}}_{\rm Q}\rightarrow \infty$ for $z\rightarrow \infty$, ${{\rm I}}_{\rm Q}=\mathrm{{\lambda}_{0}-{\lambda}_{2}}$ for $z=0$ and 
${{\rm I}}_{\rm Q}\approx \mathrm{{\lambda}_{0}-{\lambda}_{1}+{\lambda}_{2}}$ for $z\rightarrow -1$.\\
Similarly, we will focus on an interacting model with a specific ansatz for the EoS parameter, given as
\begin{equation}\label{wde}
\mathrm{\omega_{DE}}\equiv \omega_{2}+2\sum^{2}_{m=0}\frac{\omega_{m}T_{m}}{2+{z}^{2}}. 
\end{equation}
Within this ansatz a finite value for $\omega$ is obtained from the past to the future; namely, the following asymptotic values are found: 
$\mathrm{\omega_{DE}}=5\omega_{2}$ for $z\rightarrow \infty$, $\mathrm{\omega_{DE}}\approx \omega_{0}$ for $z=0$ and 
$\mathrm{\omega_{DE}}\approx(5/3)\omega_{2}+(2/3)[\omega_{0}-\omega_{1}]$ for $z\rightarrow -1$. Thus, a possible physical description 
should be explored.\\
In order to guarantee that $Q$ may be physically acceptable in the dark sectors \cite{Campo-Herrera2015}, we equal the right-hand sides of Eqs. (\ref{FormQ}) and 
(\ref{Interaction}), which becomes
\begin{equation}\label{criterion}
\frac{\mathrm{d R}}{\mathrm{d}z}=\frac{-\mathrm{R}}{(1+z)}\biggl({{\rm I}}_{\rm Q}(1+\mathrm{R})+3\mathrm{{\omega}_{DE}}\biggr). 
\end{equation}
Now, to solve or alleviate of coincidence problem, we require that $\mathrm{R}$ tends to a fixed value at late times. This leads to the condition
${\mathrm{d R}}/{\mathrm{d}z}=0$, which therefore implies two stationary solutions $\mathrm{R_{+}}=\mathrm{R}(z\rightarrow\infty)=
-(1+{3\mathrm{\omega_{DE}}}/{{{\rm I}}_{\rm Q}})$ and $\mathrm{R_{-}}=\mathrm{R}(z\rightarrow -1)=0$,
The first solution occurs in the past and the second one happens in the future.\\
By inserting Eqs. (\ref{Interaction}) and (\ref{wde}) into Eq. (\ref{criterion}), we find that $\mathrm{R}$ has no analytical solution, in any case, it is to be solved 
numerically. Likewise, there are an analytical solution for just $\mathrm{{{\rho}}_{b}}$, $\mathrm{{{\rho}}_{r}}$ and $\mathrm{{{\rho}}_{DM}}$, respectively, 
but $\mathrm{{{\rho}}_{DE}}$ will be obtained from $\mathrm{R}$, as $\mathrm{{{\rho}}_{DE}}=\mathrm{{{\rho}}_{DM}}/\mathrm{R}$.\\ 
Therefore, the first Friedmann equation is given by
\begin{equation}\label{hubble}
\smallskip
E^{2}=\frac{{\mathrm {\bf{H}}}^{2}}{\mathrm{{H}^{2}_{0}}}=\mathrm{\Omega_{b,0}}{(1+z)}^{3}+\mathrm{\Omega_{r,0}}{(1+z)}^{4}\\
+\mathrm{\Omega^{\star}_{DM}}(z)(1+{\mathrm{R}}^{-1}),
\end{equation}
where have considered that
\begin{eqnarray}
\smallskip
\mathrm{\Omega^{\star}_{DM}}(z)=(1+z)^{3}{\mathrm{\Omega_{DM,0}}}{\rm exp}\biggl[{\frac{-z_{max}}{2}\sum_{n=0}^{2}\lambda_{n}I_{n}(z)}\biggr],\qquad\qquad\qquad\nonumber\\
\int_{0}^{z}\frac{T_{n}(\tilde{x})}{(1+\tilde{x})}d\tilde{x}\,\approx\,\frac{z_{max}}{2}\int_{-1}^{x}\frac{T_{n}(\tilde{x})}{(a_{1}+a_{2}\tilde{x})}d\tilde{x}\equiv\frac{z_{max}}{2}I_{n}(z),\qquad\qquad\nonumber\\
x\,\equiv\, \frac{2z}{z_{max}}-1,\quad a_{1}\equiv1+\frac{z_{max}}{2}\,,\quad a_{2}\equiv\frac{z_{max}}{2},\qquad\qquad\qquad\nonumber\\
I_{0}(z)=\frac{2}{z_{max}}\ln(1+z),\qquad\qquad\qquad\qquad\qquad\qquad\qquad\qquad\qquad\nonumber\\
I_{1}(z)=\,\frac{2}{z_{max}}\biggl(\frac{2z}{z_{max}}-\frac{(2+z_{max})}{z_{max}}\ln(1+z)\biggr),\qquad\qquad\qquad\qquad\nonumber\\
I_{2}(z)\,=\,\frac{2}{z_{max}}\biggl[\frac{4z}{z_{max}}\biggl(\frac{z}{z_{max}}-\frac{2}{z_{max}}-2\biggr)+\qquad\qquad\qquad\qquad\quad\nonumber\\
\biggl(1+\frac{6.8284}{z_{max}}\biggr)\biggl(1+\frac{1.1716}{z_{max}}\biggr)\ln(1+z)\biggr],\nonumber\hspace{3cm}
\end{eqnarray}
where $z_{max}$ is the maximum value of $\mathrm{z}$ such that $\tilde{x}\in [-1, 1]$ and $\vert{T_{n}}(\tilde{x})\vert\leq1$ and $n\in[0,2]$ 
\cite{Cueva-Nucamendi2012}.\\ 
For a better analysis, we have compared the IDE model with other possible cosmological models. Thus, if $Q(z)=0$ and $\mathrm{\omega_{DE}}=-1$ in Eq. 
(\ref{hubble}) the standard $\Lambda$CDM model is recove-red. Similarly, when $Q(z)=0$ and $\mathrm{\omega_{DE}}$ is given by Eq. (\ref{wde}), the $\omega$DE 
model is obtained. These non-interacting models have an analytical solution for $\mathrm{R}$.
\section{Cosmographic parameters}
In this section, we are interested in studying the parameters that characterizing the kinematic state of the universe for the three models presented in 
the previous section. For this reason, we perform a Taylor series expansion of the scale factor up to the sixth order around the current epoch, $t_{0}$, 
with $\Delta{t}=t-t_{0}>0$, 
\begin{eqnarray}\label{scale1}
\mathrm{a}(t)&=&1+\mathrm{H_{0}}\Delta{t}-\frac{1}{2!}\mathrm{q_{0}}\mathrm{{H}^{2}_{0}}\Delta{t}^{2}+\frac{1}{3!}\mathrm{j_{0}H^{3}_{0}}\Delta{t}^{3}+\frac{1}{4!}\mathrm{s_{0}H^{4}_{0}}\Delta{t}^{4}\nonumber\\
&&+\frac{1}{5!}\mathrm{l_{0}{H}^{5}_{0}}\Delta{t}^{5}+\frac{1}{6!}\mathrm{m_{0}{H}^{6}_{0}}\Delta{t}^{6}+...+,
\end{eqnarray}
the coefficients of the expansion are evaluated at $t_{0}$ and allow us to define the following functions so-called cosmographics parameters of the universe 
\cite{Weinberg1972,Visser2005,Xu-Li-Lu2009,Wang-Dai-Qi2009,Vitagliano-Viel2010,Xu2011,Gruber2012,Sendra2013,Gruber2014,Movahed2017,Pan2018,Ming2017,Rodrigues2018,Ruchika2018} 
\begin{eqnarray}\label{kinematic1}
\mathrm{\bf H}&=&\frac{\dot{a}}{a}, \qquad \mathrm{\bf q}=-\frac{\ddot{a}}{a{\mathrm {\bf H}}^{2}}, \qquad \mathrm{\bf j}=-\frac{\dddot{a}}{a{{\mathrm H}}^{3}},\nonumber\\
\mathrm{\bf s}&=&\frac{\ddddot{a}}{a{\mathrm {\bf H}}^{4}}, \qquad \mathrm{\bf l}=\frac{\ddot{\dddot{a}}}{a{\mathrm{\bf H}}^{5}}, \qquad \mathrm{\bf m}=\frac{\dddot{\dddot{a}}}{a{\mathrm{\bf H}}^{6}}.
\end{eqnarray}
These functions are usually denominated as the Hubble, deceleration, jerk, snap, lerk and merk parameters, respectively. Here, the dots indicate the derivatives with 
respect to the cosmic time and without loss of generality, we have assumed that the scale factor value today, i.e., $a_{0}=1$. It is convenient to convert the 
derivatives of the above equation from time to redshift and then combine those functions among themselves, obtaining 
\begin{eqnarray}\label{kinematic2}
\mathrm{\bf{q}}&=&\frac{3\mathrm{\omega_{DE}}+1+\mathrm{R}\left(1+\frac{\mathrm{\Omega_{b,0}+2\Omega_{r,0}}}{\mathrm{\Omega_{DM,0}}}\right)}{2\biggl[1+\mathrm{R}
\left(1+\frac{\mathrm{\Omega_{b,0}+\Omega_{r,0}}}{\mathrm{\Omega_{DM,0}}}\right)\biggr]},\nonumber\\ 
\mathrm{\bf j}&=&\mathrm{\bf q}+2\mathrm{\bf q}^{2}+(1+z)\mathrm{\bf q}^{\prime},\nonumber\\
\mathrm{\bf s}&=&-\mathrm{\bf j}(2+3\mathrm{\bf q})-(1+z)\mathrm{\bf j}^{\prime},\nonumber\\
\mathrm{\bf l}&=&(1+z)\biggl[(1+z)\mathrm{\bf j}^{\prime\prime}+(3\mathrm{\bf q}^{\prime}+(6+7\mathrm{\bf q})^{\prime}\biggr]\nonumber\\
& & +\hspace{0.15cm}\mathrm{\bf j}(2+3\mathrm{\bf q})(3+4\mathrm{\bf q}),\nonumber\\
\mathrm{\bf m}&=&-(1+z)\mathrm{\bf l}^{\prime}-(4+5\mathrm{\bf q}\mathrm{\bf l}.
\end{eqnarray}
where ${}^{\prime}$ denotes the derivatives with respect to $z$. 
\section{Constraint method and observational data}
\subsection{Constraint method}
In general, to constrain the parameter spaces of the present models, we have modified the codes proposed in the MCMC method \cite{Lewis2002}. 
There are three statistical analyses that we have done to calculate the best-fit parameters: The first was done on a non-interacting model so-called 
$\Lambda$CDM with six parameters $\mathrm{P_{1}}=\mathrm{(\Omega_{DM,0},\mathrm{H_{0}},\alpha,\beta,M,dM)}$, the second was also made on a 
non-interacting scenario denominated $\omega$DE model with nine parameters 
$\mathrm{P_{2}}=\mathrm{(\omega_{0},\omega_{1},\omega_{2},\Omega_{DM,0},}\mathrm{H_{0},\alpha,\beta,M,dM)}$ and an interacting model 
with twelve parameters $\mathrm{P_{3}}=\mathrm{({\lambda}_{0},{\lambda}_{1},{\lambda}_{2},\omega_{0},}$ 
$\mathrm{\omega_{1},\omega_{2},\Omega_{DM,0},\mathrm{H_{0}},\alpha,\beta,M,dM)}$. Furthermore, the constant 
priors for the model parameters were:
$\mathrm{{\lambda}_{0}}=[-1.5\times10^{+2}+1.5\times10^{+2}]$, $\mathrm{{\lambda}_{1}}=[-1.5\times10^{+2},+1.5\times10^{+2}]$, 
$\mathrm{{\lambda}_{2}}=[-1.5\times10^{+1},+1.5\times10^{+1}]$, $\mathrm{\omega_{0}}=[-2.0,-0.3]$, $\mathrm{\omega_{1}}=[-1.0,+1.0]$, 
$\mathrm{\omega_{2}}=[-2.0,+0.1]$, $\mathrm{\Omega_{DM,0}}=[0,0.7]$, $\mathrm{H_{0}}=[20,120]$, 
$\mathrm{\alpha}=[-0.2,+0.5]$, $\mathrm{\beta}=[+2.1,+3.8]$, $\mathrm{M}=[-20,-17]$, $\mathrm{dM}=[-1.0,+1.0]$.
We have also fixed $\mathrm{\Omega_{r,0}}=\mathrm{\Omega_{\gamma,0}}(1+0.2271N_{eff})$, where $N_{eff}$ represents the effective number of neutrino 
species. So, $N_{eff}=3.04\pm0.18$, $\mathrm{\Omega_{\gamma,0}}=2.469\times10^{-5}h^{-2}$ and $\mathrm{\Omega_{b,0}}=0.02230h^{-2}$ were chosen from Table 
$4$ in \cite{Planck2015}.
\subsection{Observational data}
To test the viability of our models and set constraints on the model parameters, we use the following data sets:\\
$\bullet$ {\bf The Supernovae (SNe Ia) data}: We used the Join Analysis Luminous (JLA) \cite{Conley2011,Jonsson2010,Betoule2014} data 
composed by $740$ SNe Ia with hight-quality light curves, which include samples from $z<0.1$ to $0.2<z<1.0$.\\
The observed distance modulus is modeled by \cite{Conley2011,Jonsson2010,Betoule2014}
\begin{equation}\label{muJLA}
{\mu}^{JLA}_{i}={m}^{*}_{B,i}+\mathrm{\alpha}{x_{1,i}}-\mathrm{\beta}C_{i}-\mathrm{M}-\mathrm{dM},\quad 1\leq i\leq740,
\end{equation}
where and the parameters ${m}^{*}_{B}$, $x_{1}$ and $C$ describe the intrinsic variability in the luminosity of the SNe. Furthermore, the nuisance 
parameters $\mathrm{\alpha}$, $\mathrm{\beta}$, $\mathrm{M}$ and $\mathrm{dM}$ characterize the global pro-perties of the light-curves of the SNe and are 
estimated simultaneously with the cosmological parameters of interest. Then, the theoretical distance modulus is 
\begin{equation}\label{mus}
{\mu}^{\rm{th}}(z) \equiv 5{\log}_{10}\left[\frac{{D_{L}}(z)}{\rm{Mpc}}\right]+25,
\end{equation}
where ``$\rm{th}$'' denotes the theoretical prediction for a SNe at $z$. The luminosity distance ${D_{L}}(z)$, is defined as
\begin{equation}\label{luminosity_distance1}
{D}_{L}(z_{hel},z_{CMB})=(1+z_{hel})c \int_{0}^{z_{CMB}}\frac{dz'}{\mathrm{\bf H}(z')},
\end{equation}
where $z_{hel}$ is the heliocentric redshift, $z_{CMB}$ is the CMB rest-frame redshift, $c=2.9999\times10^{5}km/s$ is the speed of the light. Thus,
\begin{eqnarray}\label{mus}
{\mu}^{{\rm th}}(z_{hel},z_{CMB})&=&5\log_{10}\biggl[(1+z_{hel}\int_{0}^{z_{CMB}}\frac{dz'}{E(z'})\biggr]\nonumber\\
&&+52.385606-5\log_{10}(\mathrm{H_{0}}).
\end{eqnarray}
Then, the ${\chi}^{2}$ distribution function for the JLA data is
\begin{equation}\label{X2JLA}
{\chi}_{\bf JLA}^{2}=\left({\Delta{\mu}}_{i}\right)^{t}\left(C^{-1}_{\bf Betoule}\right)_{ij}\left({\Delta{\mu}}_{j}\right),
\end{equation}
where ${\Delta \mu}_{i}={\mu}^{th}_{i}-{\mu}^{JLA}_{i}$ is a column vector and $C^{-1}_{\bf Betoule}$ is the $740\times740$ covariance 
matrix \cite{Betoule2014}.\\\\
$\bullet$ {\bf Baryon Acoustic Oscillation (BAO) data}: The BAO distance measurements can be used to constrain the distance ratio
${d_{z}(z)}=\frac{r_{s}(z_{d})}{D_{V}(z)}$ at different redshifts, obtained from different surveys 
\cite{Hinshaw2013,Beutler2011,Ross2015,Percival2010,Blake2011,Kazin2010,Anderson2014a,Padmanabhan2012,Chuang2013a,Chuang2013b,Debulac2015,FontRibera2014}
listed in Table \ref{tableBAOI}. Here, $r_{s}({z_{d}})$ is the comoving sound horizon size at the baryon drag epoch ${z_{d}}$, 
where the baryons were released from photons and has been calculated by \cite{Eisenstein1998}. Moreover, the dilation scale is defined as 
$D_{v}(z)\equiv\frac{1}{\mathrm{H_{0}}}\left[(1+z)^{2}{D_{A}}^{2}(z)\frac{cz}{E(z)}\right]^{1/3}$, 
where $D_{A}(z)=c{\int}^{z}_{0}\frac{dz'}{\mathrm{\bf H}(z')}$ is the angular diameter distance. Thus, the $\chi^{2}$ is given as
\begin{equation}\label{X2BAOI}
\chi_{\bf{{BAO}}\,\rm{I}}^{2}=\sum_{i=1}^{17}\left(\frac{d_{z}^{\mathrm{th}}(\mathrm{z_{i}})-d_{z}^{\mathrm{obs}}(\mathrm{z_{i}})}{\sigma(\mathrm{z_{i}})}\right)^{2}.
\end{equation}
\begingroup
\begin{table}[!htb]
\centering
\resizebox{0.45\textwidth}{!}{
\begin{tabular}{*{8}{|l}|}
\hline\noalign{\smallskip}
 $z$ & ${d_{z}^{obs}}$ & $\sigma_{z}$& Refs. & $z$ & ${d_{z}^{obs}}$ & $\sigma$& Refs.\\
\noalign{\smallskip}\hline\noalign{\smallskip}
$0.106$ & $0.3360$ &$\pm0.0150$ &\cite{Hinshaw2013,Beutler2011} &$0.350$ & $0.1161$ &$\pm0.0146$&\cite{Chuang2013a}\\
$0.150$ & $0.2232$ &$\pm0.0084$ &\cite{Ross2015} &$0.440$ & $0.0916$ &$\pm0.0071$&\cite{Blake2011}\\
$0.200$ & $0.1905$ &$\pm0.0061$ &\cite{Percival2010,Blake2011} &$0.570$ & $0.0739$ &$\pm0.0043$&\cite{Chuang2013b}\\
$0.275$ & $0.1390$ &$\pm0.0037$ &\cite{Percival2010} &$0.570$ & $0.0726$ &$\pm0.0014$&\cite{Anderson2014a}\\
$0.278$ & $0.1394$ &$\pm0.0049$ &\cite{Kazin2010} &$0.600$ & $0.0726$ &$\pm0.0034$&\cite{Blake2011}\\
$0.314$ & $0.1239$ &$\pm0.0033$ &\cite{Blake2011} &$0.730$ & $0.0592$ &$\pm0.0032$&\cite{Blake2011}\\
$0.320$ & $0.1181$ &$\pm0.0026$ &\cite{Anderson2014a}&$2.340$ & $0.0320$ &$\pm0.0021$&\cite{Debulac2015}\\
$0.350$ & $0.1097$ &$\pm0.0036$ &\cite{Percival2010,Blake2011} &$2.360$ & $0.0329$ &$\pm0.0017$&\cite{FontRibera2014}\\
$0.350$ & $0.1126$ &$\pm0.0022$ &\cite{Padmanabhan2012}        &$ $&$ $\\ 
\noalign{\smallskip}\hline
\end{tabular}}
\caption{Summary of BAO\,I data 
\cite{Hinshaw2013,Beutler2011,Ross2015,Percival2010,Blake2011,Kazin2010,Anderson2014a,Padmanabhan2012,Chuang2013a,Chuang2013b,Debulac2015,FontRibera2014}.}\label{tableBAOI}
\end{table}
\endgroup
\\
\begingroup
\begin{table}[!htb]
\centering
\resizebox{0.45\textwidth}{!}{
\begin{tabular}{*{8}{|l}|}
\hline\noalign{\smallskip}
 $z$   & ${\mathrm{\bf H}}(z)$ &  $1\sigma$& Refs. & $\mathrm{z}$   & ${\mathrm{\bf H}}(z)$ &  $1\sigma$& Refs. \\
\noalign{\smallskip}\hline\noalign{\smallskip}
$0.070$&  $69.0$&  $\pm19.6$&\cite{Zhang2014}    & $0.480$&  $97.0$&  $\pm62.0$&\cite{Stern2010}\\  
$0.090$&  $69.0$&  $\pm12.0$&\cite{Simon2005}    & $0.570$&  $87.6$&  $\pm7.80$&\cite{Chuang2013b}\\ 
$0.120$&  $68.6$&  $\pm26.2$&\cite{Zhang2014}    & $0.570$&  $96.8$&  $\pm3.40$&\cite{Anderson2014a}\\ 
$0.170$&  $83.0$&  $\pm8.0$&\cite{Simon2005}     & $0.593$& $104.0$&  $\pm13.0$&\cite{Moresco2012}\\
$0.179$&  $75.0$&  $\pm4.0$&\cite{Moresco2012}   & $0.600$&  $87.9$&  $\pm6.1$ &\cite{Blake2012}\\ 
$0.199$&  $75.0$&  $\pm5.0$&\cite{Moresco2012}   & $0.680$&  $92.0$&  $\pm8.0$ &\cite{Moresco2012}\\
$0.200$&  $72.9$&  $\pm29.6$&\cite{Zhang2014}    & $0.730$&  $97.3$&  $\pm7.0$ &\cite{Blake2012}\\
$0.240$&  $79.69$& $\pm2.99$&\cite{Gastanaga2009}& $0.781$& $105.0$&  $\pm12.0$&\cite{Moresco2012}\\
$0.270$&  $77.0$&  $\pm14.0$&\cite{Simon2005}    & $0.875$& $125.0$&  $\pm17.0$&\cite{Moresco2012}\\
$0.280$&  $88.8$&  $\pm36.6$&\cite{Zhang2014}    & $0.880$&  $90.0$&  $\pm40.0$&\cite{Stern2010}\\
$0.300$&  $81.7$&  $\pm6.22$&\cite{Oka2014}      & $0.900$& $117.0$&  $\pm23.0$&\cite{Simon2005}\\
$0.340$&  $83.8$&  $\pm3.66$&\cite{Gastanaga2009}& $1.037$& $154.0$&  $\pm20.0$&\cite{Gastanaga2009}\\
$0.350$&  $82.7$&  $\pm9.1$& \cite{Chuang2013a}  & $1.300$& $168.0$&  $\pm17.0$&\cite{Simon2005}\\
$0.352$&  $83.0$&  $\pm14.0$&\cite{Moresco2012}  & $1.363$& $160.0$&  $\pm33.6$&\cite{Moresco2015}\\
$0.3802$& $83.0$&  $\pm13.5$&\cite{Moresco2016}  & $1.430$& $177.0$&  $\pm18.0$&\cite{Simon2005}\\
$0.400$&  $95.0$&  $\pm17.0$&\cite{Simon2005}    & $1.530$& $140.0$&  $\pm14.0$&\cite{Simon2005}\\
$0.4247$& $87.1$&  $\pm11.2$&\cite{Moresco2016}  & $1.750$& $202.0$&  $\pm40.0$&\cite{Simon2005}\\
$0.430$&  $86.45$& $\pm3.97$&\cite{Gastanaga2009}& $1.965$& $186.5$&  $\pm50.4$&\cite{Moresco2015}\\
$0.440$&  $82.6$&  $\pm7.8$&\cite{Blake2012}     & $2.300$& $224.0$&  $\pm8.6$ &\cite{Busca2013}\\ 
$0.4497$& $92.8$&  $\pm12.9$&\cite{Moresco2016}  & $2.340$& $222.0$&  $\pm8.5$ &\cite{Debulac2015}\\
$0.4783$& $80.9$&  $\pm9.0$&\cite{Moresco2016}   & $2.360$& $226.0$&  $\pm9.3$ &\cite{FontRibera2014}\\
\noalign{\smallskip}\hline
\end{tabular}}
\caption{Shows the ${\mathrm{\bf H}(z)}$ data 
\cite{Anderson2014a,Chuang2013a,Chuang2013b,Debulac2015,FontRibera2014,Zhang2014,Simon2005,Moresco2012,Moresco2016,Gastanaga2009,Oka2014,Blake2012,Stern2010,Moresco2015,Busca2013}}\label{tableOHD} 
\end{table}
$\bullet$ {\bf Cosmic Microwave Backgroung data}: We use the Planck distance priors data extracted from Planck $2015$ results XIII Cosmological parameters, 
for the combined analysis TT, TF, FF + lowP + lensing \cite{Planck2015,Neveu2016}. From here, we have obtained the values of the shift parameter ${R(z_{*})}$, 
the angular scale for the sound horizon at photon-decoupling epoch, $l_{A}(z_{*})$, and the redshift at photon-decoupling epoch, $z_{*}$.
Then, the shift parameter $\tilde{R}$ is defined by \cite{Bond-Tegmark1997}
\begin{equation}\label{Shiftparameter}
\tilde{R}(z_{*})\equiv\sqrt{\Omega_{M,0}}{\int}^{z_{*}}_{0}\frac{d\tilde{y}}{E(\tilde{y})},
\end{equation}
where $E(\tilde{y})$ is given by Eq. (\ref{hubble}) and the redshift $z_{*}$ is obtained from \cite{Hu-Sugiyama1996}
\begin{equation}\label{Redshift_decoupling}
{z}_{*}=1048\biggl[1+0.00124(\mathrm{{\Omega}_{b,0}}h^{2})^{-0.738}\biggr]\biggl[1+{g}_{1}(\mathrm{{\Omega}_{M,0}}h^{2})^{{g}_{2}}\biggr]\;,\;\;
\end{equation}
where 
\begin{equation}\label{g1g2}
g_{1}=\frac{0.0783(\mathrm{\Omega_{b,0}}h^{2})^{-0.238}}{1+39.5(\mathrm{\Omega_{b,0}}h^{2})^{0.763}},\hspace{0.3cm}
g_{2}=\frac{0.560}{1+21.1(\mathrm{\Omega_{b,0}}h^{2})^{1.81}}.
\end{equation}
The angular scale $l_{A}$ for the sound horizon is 
\begin{equation}\label{Acoustic_scale}
l_{A}\equiv\frac{\pi D_{A}(z_{*})}{r_{s}(z_{*})},\hspace{1cm}
\end{equation}
where $r_{s}(z_{*})$ is the comoving sound horizon at $z_{*}$. From \cite{Planck2015,Neveu2016}, the $\chi^{2}$ is
\begin{equation}\label{X2CMB}
\chi_{\bf CMB}^{2}=\left({\Delta x}_{i}\right)^{t}\left(C^{-1}_{\bf CMB}\right)_{ij}\left({\Delta x}_{j}\right),
\end{equation}
where ${\Delta x}_{i}=x^{\mathrm{th}}_{i}-x^{\mathrm{obs}}_{i}$ is a column vector 
\begin{equation}
x^{\mathrm{th}}_{i}(\mathbf{X})-x^{\mathrm{obs}}_{i}=\left(\begin{array}{cc}
 l_{A}(z_{*})-301.7870\\
 R(z_{*})-1.7492\\
 \;\;z_{*}-1089.990\\
\end{array}\right),
\end{equation}
``t'' denotes its transpose and $(C^{-1}_{\bf CMB})_{ij}$ is the inverse covariance matrix \cite{Neveu2016}
given by
\begin{equation}\label{MatrixCMB}
C^{-1}_{\bf CMB}\equiv\left(
\begin{array}{ccc}
+162.48&-1529.4&+2.0688\\
-1529.4&+207232&-2866.8\\
+2.0688&-2866.8&+53.572\\
\end{array}\right).
\end{equation}
$\bullet$ {\bf Hubble observational data}:
This sample is composed by 42 independent measurements of the Hubble parameter at different redshifts and were derived 
from differential age $dt$ for passively evolving galaxies with redshift $dz$ and from the two-points correlation function of Sloan Digital Sky Survey. 
This sample was taken from Table $\rm{III}$ in \cite{Sharov2015,Moresco2016}. Then, the $\chi^2_{\mathrm{\bf H}}$ function for this data set is \cite{Sharov2015}
\begin{equation}\label{X2OHD}
\chi^2_{\bf{{H}}}\equiv\sum_{i=1}^{42}\frac{\left[{\mathrm H}^{\rm {th}}(z_{i})-{\mathrm H}^{\mathrm{obs}}(\mathrm{z_{i}})\right]^2}{\sigma^2(z_{i})},
\end{equation}
where ${\mathrm H}^{{\rm th}}$ denotes the theoretical value of $\mathrm{\bf H}$, ${\mathrm H}^{\mathrm{obs}}$ represents its observed value and 
$\sigma(z_{i})$ is the error.\\
In order to put constraints on the model parameters, we have calculated the overall likelihood $\mathcal{L}\, \alpha\, {{\mathrm e}}^{-{{\chi}}^{2}/2}$, 
where ${\rm {\bf{\chi}}^{2}}$ can be defined by
\begin{equation}\label{TotalChi}
{{{\rm {\bf{\chi}}^{2}}}}={{{\rm{\bf{\chi}}^{2}}}}_{\bf JLA}+{{{\rm {\bf {\chi}}^{2}}}}_{\bf BAO}+{{{\rm {\bf {\chi}}^{2}}}}_{\bf CMB}+
{{{\rm {\bf {\chi}}^{2}}}}_{\bf H}\,.
\end{equation}
\begin{table*}[!hbtp]
\centering
\resizebox{0.9\textwidth}{!}{
\begin{tabular}{*{6}{|c}|}
\hline\noalign{\smallskip}
Parameters&$\Lambda$CDM&$\omega$DE&IDE1&IDE2\\
\noalign{\smallskip}\hline\noalign{\smallskip} 
$\mathrm{{\lambda}_{0}}\times10^{+4}$&$N/A$&$N/A$&${+1.120}^{+0.6569+2.1166}_{-0.5709-1.4348}$&${+1.120}^{+0.6569+2.1166}_{-0.5709-1.4348}$\\[0.4mm]
$\mathrm{{\lambda}_{1}}\times10^{+4}$&$N/A$&$N/A$&${+2.733}^{+0.4108+0.7581}_{-0.5648-1.9464}$&${+2.733}^{+0.4108+0.7581}_{-0.5648-1.9464}$\\[0.4mm]
$\mathrm{{\lambda}_{2}}\times10^{+5}$&$N/A$&$N/A$&${+2.7112}^{+0.8670+1.6127}_{-3.5277-4.6376}$&${-2.6490}^{+0.5032+1.0396}_{-1.1769-2.8069}$\\[0.6mm]
$\mathrm{\omega_{0}}$&$-1.0$&${-1.0364}^{+0.0648+0.1140}_{-0.0865-0.1910}$&${-1.0730}^{+0.0644+0.1139}_{-0.0863-0.1906}$&${-1.0773}^{+0.0645+0.1136}_{-0.0863-0.1918}$\\[0.4mm]
$\mathrm{\omega_{1}}$&$N/A$&${+1.150}^{+0.2377+0.5849}_{-0.1230-0.1965}$&${+1.2941}^{+0.2377+0.5848}_{-0.1230-0.1965}$&${+1.2969}^{+0.2359+0.5849}_{-0.1239-0.1964}$\\[0.4mm]
$\mathrm{\omega_{2}}$&$N/A$&${-1.0546}^{+0.1283+0.4803}_{-0.0381-0.0741}$&${-0.6179}^{+0.1281+0.4802}_{-0.0382-0.0742}$&${-0.6179}^{+0.1281+0.4802}_{-0.0382-0.0742}$\\[0.4mm]
$\mathrm{\Omega_{DM,0}}$&${+0.2812}^{+0.0185+0.0478}_{-0.0140-0.0278}$&${+0.2844}^{+0.0118+0.0386}_{-0.0062-0.0125}$&${+0.2844}^{+0.0118+0.0386}_{-0.0062-0.0125}$
&${+0.2844}^{+0.0118+0.0386}_{-0.0062-0.0125}$\\[0.4mm]
$\mathrm{\Omega_{b,0}}$&${+0.0493}^{+0.0018+0.0037}_{-0.0020-0.0040}$&${+0.0494}^{+0.0012+0.0025}_{-0.0014-0.0030}$&${+0.0494}^{+0.0012+0.0025}_{-0.0014-0.0030}$&${+0.0494}^{+0.0012+0.0025}_{-0.0014-0.0030}$\\[0.4mm]
$\mathrm{H_{0}}$&${+67.10}^{+1.3038+2.6882}_{-1.3336-2.5704}$&${+67.1480}^{+0.8092+1.8055}_{-0.9875-1.9252}$&${+67.1487}^{+0.8085+1.8048}_{-0.9882-1.9259}$
&${+67.1487}^{+0.8085+1.8048}_{-0.9882-1.9259}$\\[0.4mm]
$\mathrm{\alpha}$&${+0.1360}^{+0.0418+0.0848}_{-0.0403-0.0810}$&${+0.1360}^{+0.1103+0.2289}_{-0.1205-0.2483}$&${+0.1360}^{+0.1103+0.2289}_{-0.1205-0.2483}$&${+0.1360}^{+0.1103+0.2289}_{-0.1205-0.2483}$\\[0.4mm]
$\mathrm{\beta}$&${+3.060}^{+0.1031+0.2083}_{-0.1058-0.2069}$&${+3.0780}^{+0.1964+0.3917}_{-0.1858-0.3708}$&${+3.0780}^{+0.1964+0.3917}_{-0.1858-0.3708}$&${+3.0780}^{+0.1964+0.3917}_{-0.1858-0.3708}$\\[0.4mm]
$\mathrm{M}$&${-19.0324}^{+0.3796+0.7651}_{-0.3888-0.7769}$&${-19.0880}^{+0.5637+1.1179}_{-0.5527-1.1042}$&${-19.0631}^{+0.5647+1.1099}_{-0.5517-1.0970}$&${-19.0631}^{+0.5647+1.1099}_{-0.5517-1.0970}$\\[0.4mm]
$\mathrm{dM}$&${-0.124}^{+0.2774+0.5409}_{-0.2721-0.5338}$&${-0.1230}^{+0.3711+0.7483}_{-0.3832-0.7503}$&${-0.1230}^{+0.3711+0.7483}_{-0.3832-0.7503}$&${-0.1230}^{+0.3711+0.7483}_{-0.3832-0.7503}$\\[0.4mm]
\hline${\chi}^{2}_{min}$&$719.8072$&$707.8162$&$704.0342$&$703.2585$\\[0.4mm]
\noalign{\smallskip}\hline
\end{tabular}}
\caption{Shows the best-fit values of the cosmological parameters for the three models with $1\sigma$ and $2\sigma$ errors.}\label{Bestfits}
\end{table*}
\begin{table*}[!hbtp]
\centering
\resizebox{0.7\textwidth}{!}{
\begin{tabular}{*{6}{|c}|}
\hline\noalign{\smallskip}
Parameters&$\Lambda$CDM&$\omega$DE&IDE1&IDE2\\
\noalign{\smallskip}\hline\noalign{\smallskip} 
$\mathrm{q_{0}}$&${-0.5042}^{+0.0253}_{-0.0176}$&${-0.5357}^{+0.0805}_{-0.0936}$&${-0.5721}^{+0.0805}_{-0.0945}$&${-0.5764}^{+0.0806}_{-0.0944}$\\[0.4mm]
$\mathrm{j_{0}}$&${+1.0}$&${+2.2625}^{+0.1854}_{-0.0205}$&${+2.5280}^{+0.2082}_{-0.0037}$&${+2.5458}^{+0.2097}_{-0.0001}$\\[0.4mm]
$\mathrm{s_{0}}$&${-0.4880}^{+0.0529}_{-0.0758}$&${+7.2042}^{+1.3190}_{-1.2697}$&${+6.0919}^{+1.5122}_{-1.3192}$&${+6.1821}^{+1.5200}_{-1.3264}$\\[0.4mm]
$\mathrm{l_{0}}$&${+3.4698}^{+0.2047}_{-0.1383}$&${+22.780}^{+8.1174}_{-1.4726}$&${+27.5037}^{+6.4060}_{-0.2759}$&${+28.0020}^{+6.4938}_{-0.1259}$\\[0.4mm]
$\mathrm{m_{0}}$&${-17.6637}^{+1.1241}_{-1.6560}$&${+190.5955}^{+24.3474}_{-19.4601}$&${+133.6285}^{+41.3932}_{-43.7943}$&${+136.2745}^{+42.0971}_{-54.1525}$\\[0.4mm]
${z_{t}}$&${+0.595}^{+0.030}_{-0.040}$&${+0.580}^{+0.270}_{-0.255}$&${+0.540}^{+1.165}_{-0.220}$&${0.545}^{+1.160}_{-0.215}$\\[0.4mm]
\noalign{\smallskip}\hline
\end{tabular}}
\caption{Shows the current values of the cosmographic parameters and the transition redshift $z_{t}$ with $1\sigma$ error.}\label{Bestfitstoday}
\end{table*}
\section{Results.} \label{SectionAllTest}
In this work, we have run eight chains for each of the three models proposed on the computer, and the best-fit 
parameters with $1\sigma$ and $2\sigma$ errors, are presented in Table \ref{Bestfits}. Hence, we can see that the corresponding 
${\chi}^{2}_{min}$ for the IDE model becomes smaller in comparison with those obtained in the non-interacting models.\\ 
The one-dimension probability contours with $1\sigma$ and $2\sigma$ errors on each parameter of the 
present models and obtained from the combined constraint of geometric data, are plotted in Fig. \ref{Contours}.\\
Due to the two minimums obtained in the IDE model (see Table \ref{Bestfits}), we consider now two different cases to reconstruct ${\rm I}_{\rm Q}$: 
the case 1 is so-called IDE1 with ${\lambda}_{2}>0$; by contrast, the case 2 is dubbed IDE2 with ${\lambda}_{2}<0$.\\
The evolution of $\mathrm{\omega_{DE}}$ with respect to redshift and within the $1\sigma$ error around the best-fit curve 
for the present models, is presented in the left upper panel of Fig. \ref{Reconstructions1}. From here, one can see that in the $\mathrm{\omega}$DE 
and IDE models, the universe evolves from the phantom regime $\mathrm{\omega_{DE}} < -1$ to the quintessence regime $\mathrm{\omega_{DE}} > -1$, and then 
it becomes phantom again. Moreover, $\mathrm{\omega_{DE}}$ crosses the phantom divide line $-1$ \cite{Nesseris2007} twice. In particular, the IDE1 (IDE2) 
case has two crossing points in the $1\sigma$ confidence region in $z={+0.0591}^{+0.1024}_{-0.0536}$ 
($z={+0.0633}^{+0.1038}_{-0.0555}$) and $z={+1.1794}^{+0.9290}_{-0.3137}$ ($z={+1.1784}^{+0.9288}_{-0.3163}$), 
respectively. Analogously, for the $\omega$DE model these points are respectively $z={+0.0337}^{+0.1970}_{-0.0535}$ and 
$z={+0.5024}^{+0.2827}_{-0.2730}$. Such a crossing feature is favored by the data within $1\sigma$ error.\\
Likewise, our fitting results show that the evolution of $\mathrm{\omega_{DE}}$ in the $\omega$DE and IDE models are very close to each other, in particular, 
they are close to $-1$ today. These results imply that $\mathrm{\omega_{DE}}$ shows a phantom nature today and are in excellent agreement with the 
constraints at $1\sigma$ confidence region obtained by \cite{Planck2015}.\\ 
The evolution of ${\rm I}_{\rm Q}$ along $z$ and within the $1\sigma$ error around the best-fit curve 
for the IDE model is shown in the right upper panel of Fig. \ref{Reconstructions1}. From where, we see that ${\rm I}_{\rm Q}$ can change 
its sign throughout its evolution. Now, from Eqs. (\ref{EDM}) and (\ref{EDE}), we conveniently establish the following convention: ${\rm I}_{+}$ denotes 
an energy transfer from DE to DM while ${\rm I}_{-}$ denotes an energy transfer from DM to DE. From here, we have found a change from 
${\rm I}_{+}$ to ${\rm I}_{-}$ and vice versa. This change of sign is linked to the crossing of the line, ${\rm I}_{\rm Q}=0$, which is also 
favored by the data at $1\sigma$ error. The IDE model shows three crossing points in $z={-0.3370}^{+0.0553}_{-0.1734}$ (IDE1), 
$z={-0.4673}^{+0.0874}_{-0.1181}$ (IDE2) and $z={+5.6077}^{+2.2901}_{-2.3950}$ (IDE2), respectively.\\
The fitting results indicate that ${\rm I}_{\rm Q}$ is stronger at early times and weaker at later 
times, namely, ${\rm I}_{\rm Q}$ remains small today, being ${\rm I}_{\rm Q,0}={+8.5875\times10^{-5}}^{+5.6690\times10^{-5}}_{-2.2592\times10^{-5}}$ 
for the case IDE1 and ${\rm I}_{\rm Q,0}={+13.8237\times10^{-5}}^{+6.0544\times10^{-5}}_{-4.5652\times10^{-5}}$ for the case IDE2, respectively. 
These results are consistent at $1\sigma$ error with those reported in \cite{Cueva-Nucamendi2012,Cai2010,Li2011}. However, our outcomes are smaller with 
tighter constraints. This discrepancy may be due to the ansatz chosen for ${\rm I}_{\rm Q}$ and the used data.\\
For the three cosmologies, the background expansion rate $\mathrm{{\bf H}/H_{0}}$ with respect to $z$ is shown in left below panel of the Fig. 
\ref{Reconstructions1}. To emphasize a possible deviation at $z<0$, we have plotted up to $z=-1.0$. Hence, we have noted 
that the amplitudes of $\mathrm{{\bf H}/H_{0}}$ in the $\omega$DE and IDE models deviate significantly from that found in the $\Lambda$CDM model. 
It means that $\mathrm{{\bf H}/H_{0}}$ is sensitive with both $Q$ and $\mathrm{\omega_{DE}}$. 
\begin{figure*}[!htb]
\centering
\resizebox{0.95\textwidth}{!}{
\begin{tabular}{*{2}{ c }}
\includegraphics{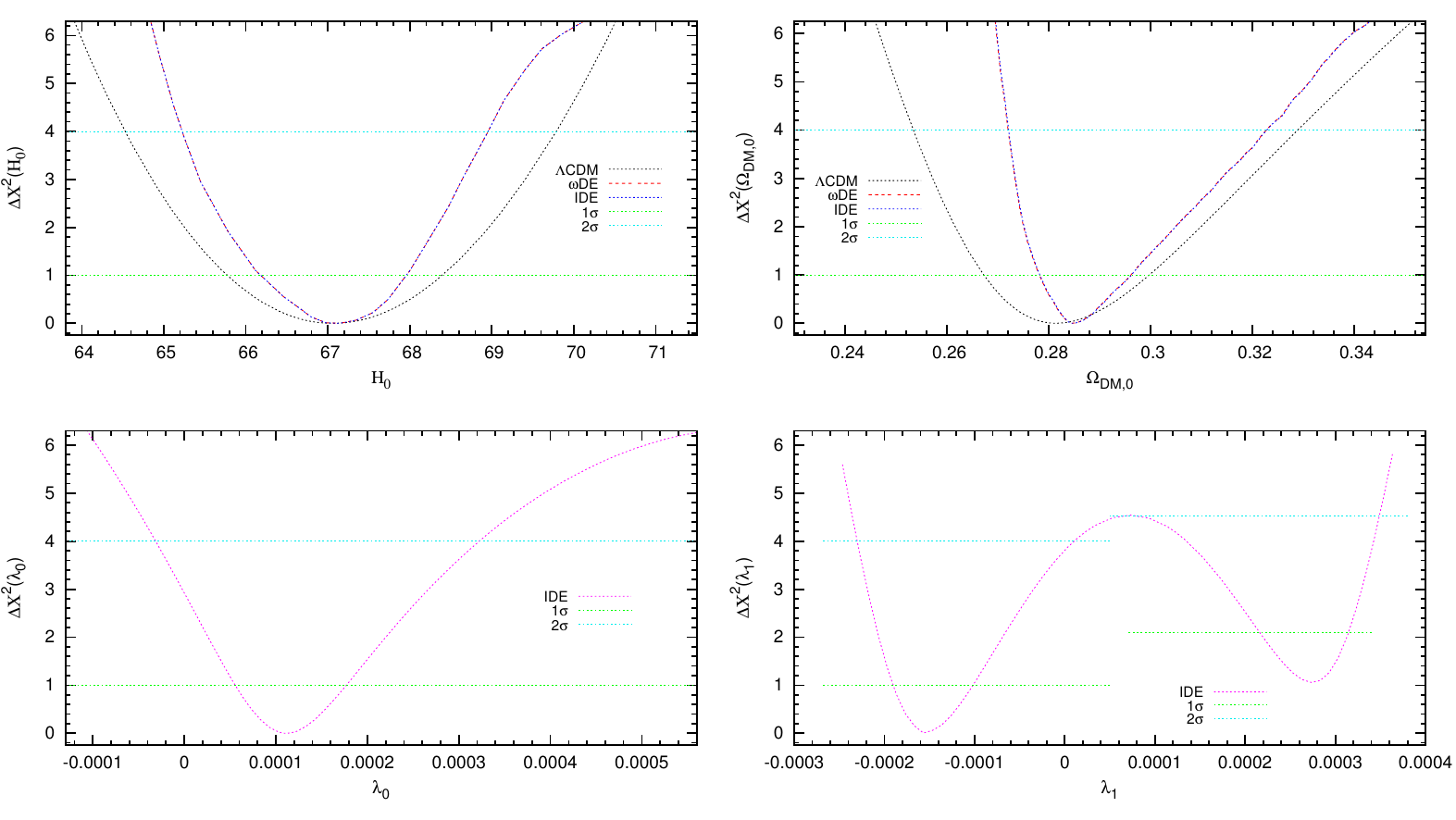}&\includegraphics{B1.pdf}\\
\includegraphics{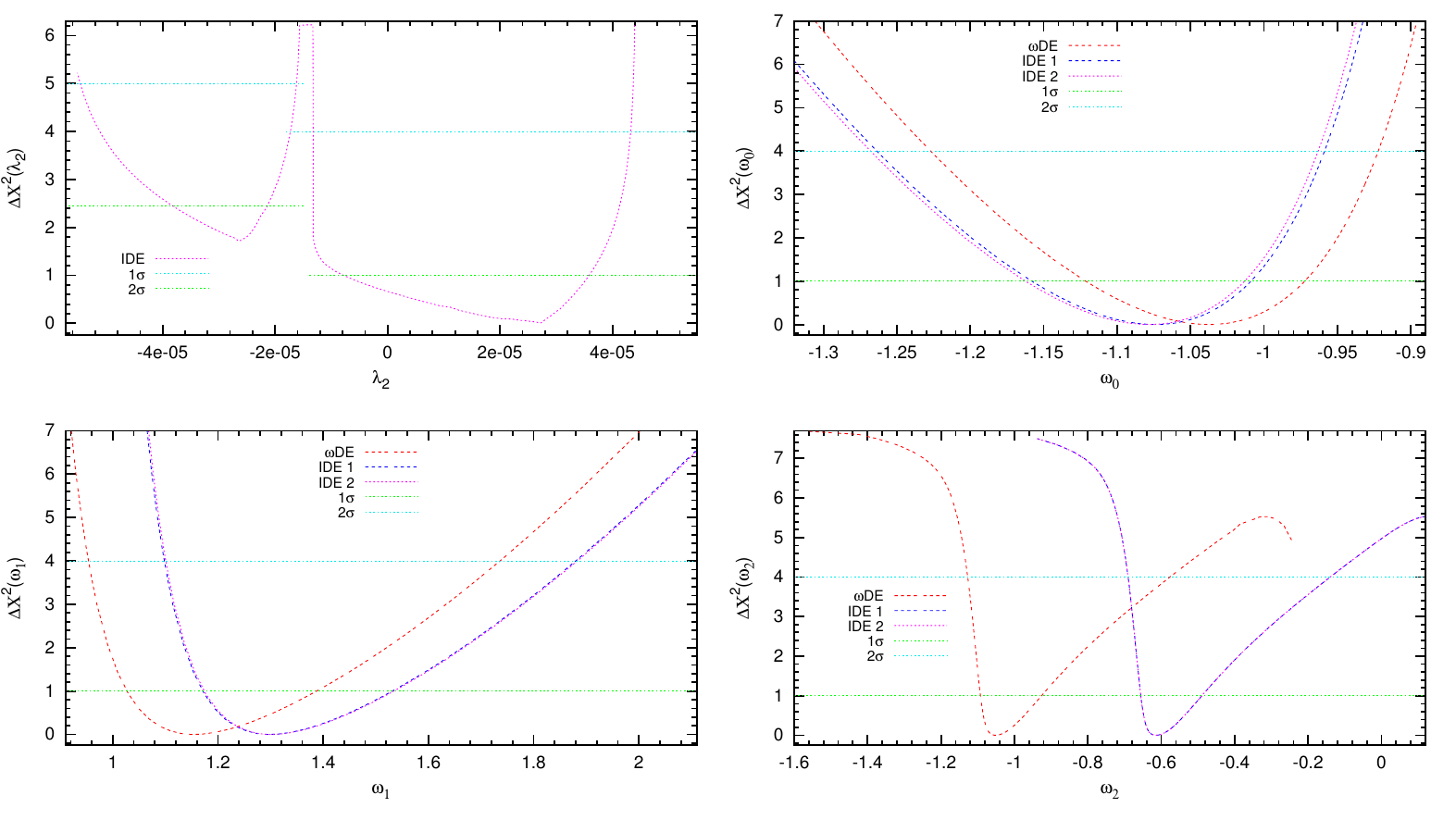}
\end{tabular}} 
\caption{\label{Contours} Displays the one-dimension probability contours of the parameter space at $1\sigma$ and 
$2\sigma$ errors. Besides $\mathrm{\Delta{\chi^{2}}=\chi^{2}-{\chi^{2}_{min}}}$.}
\end{figure*}
\begin{figure*}[!htb]
 \centering 
 \resizebox{0.75\textwidth}{!}{
 \begin{tabular}{*{2}{ c }}
 \includegraphics{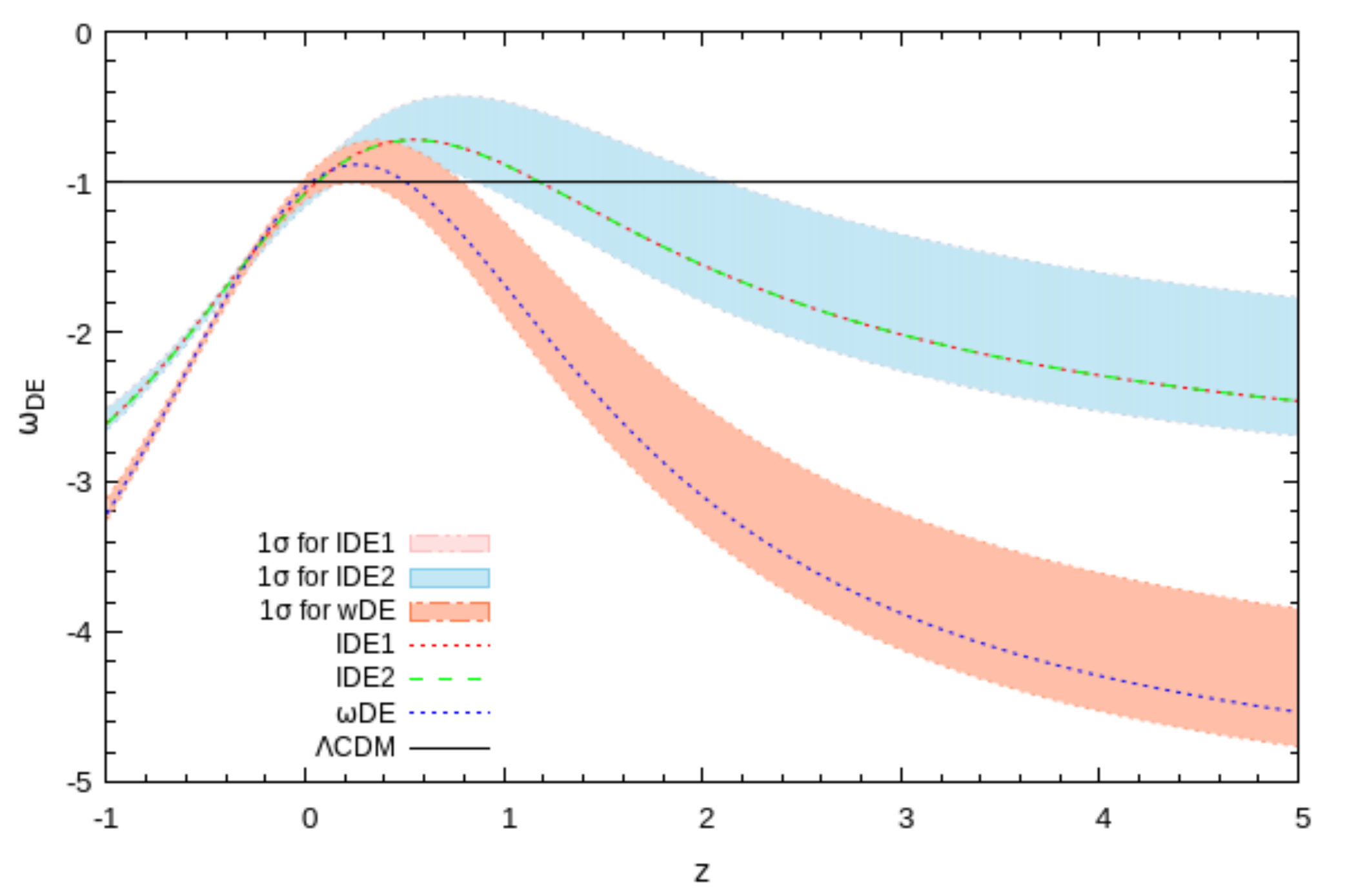}&\includegraphics{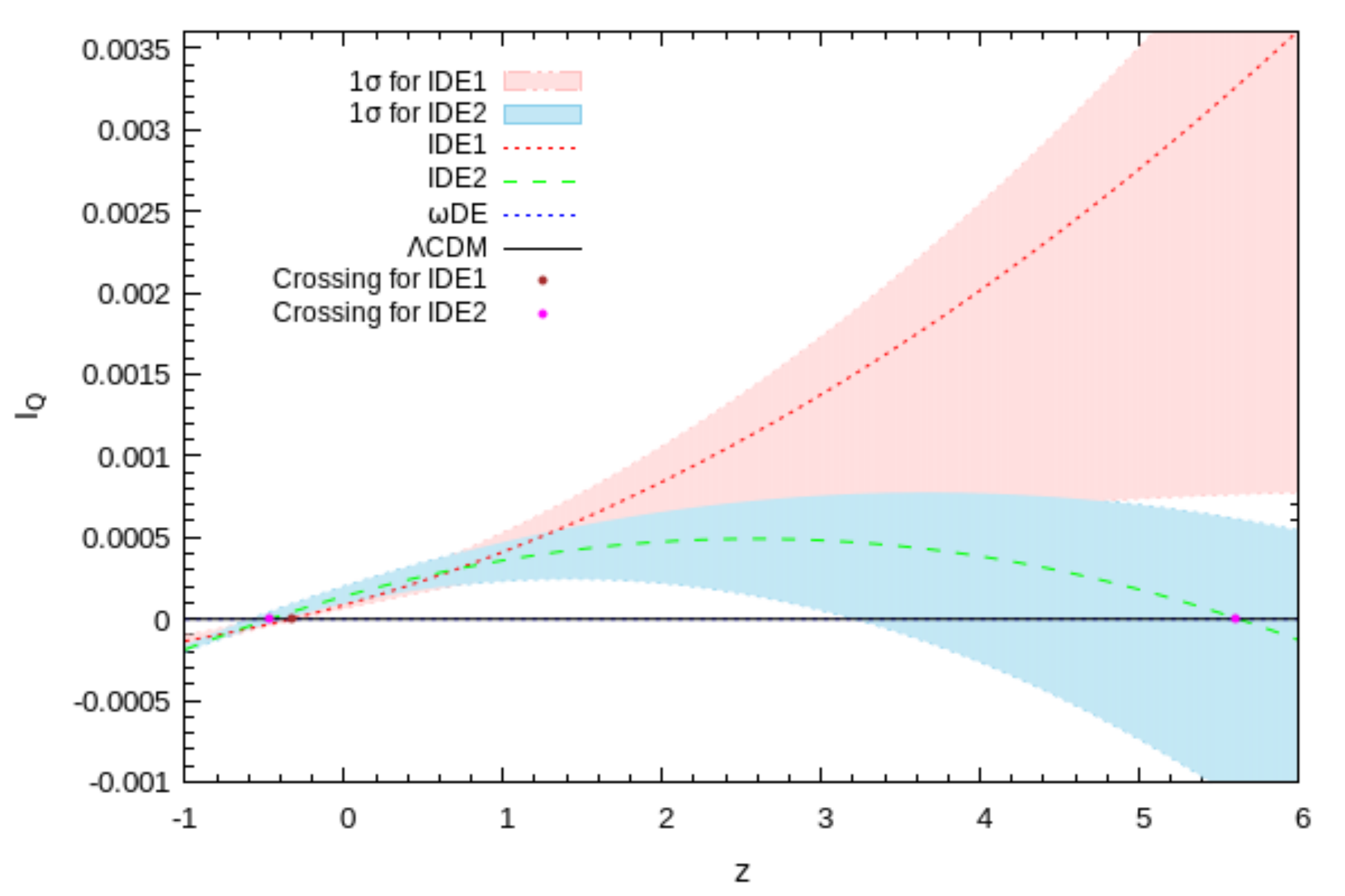}\\
 \includegraphics{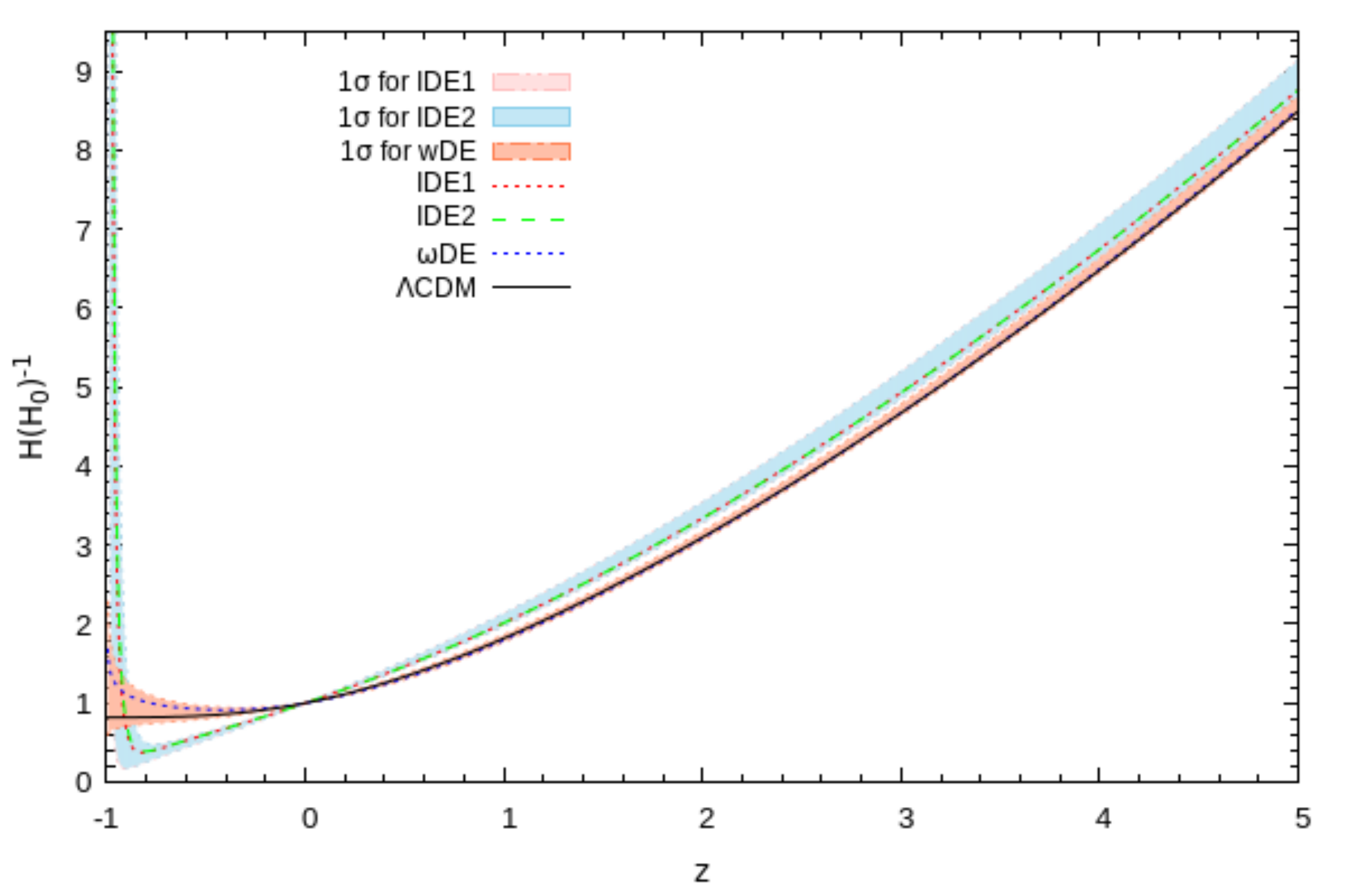}&\includegraphics{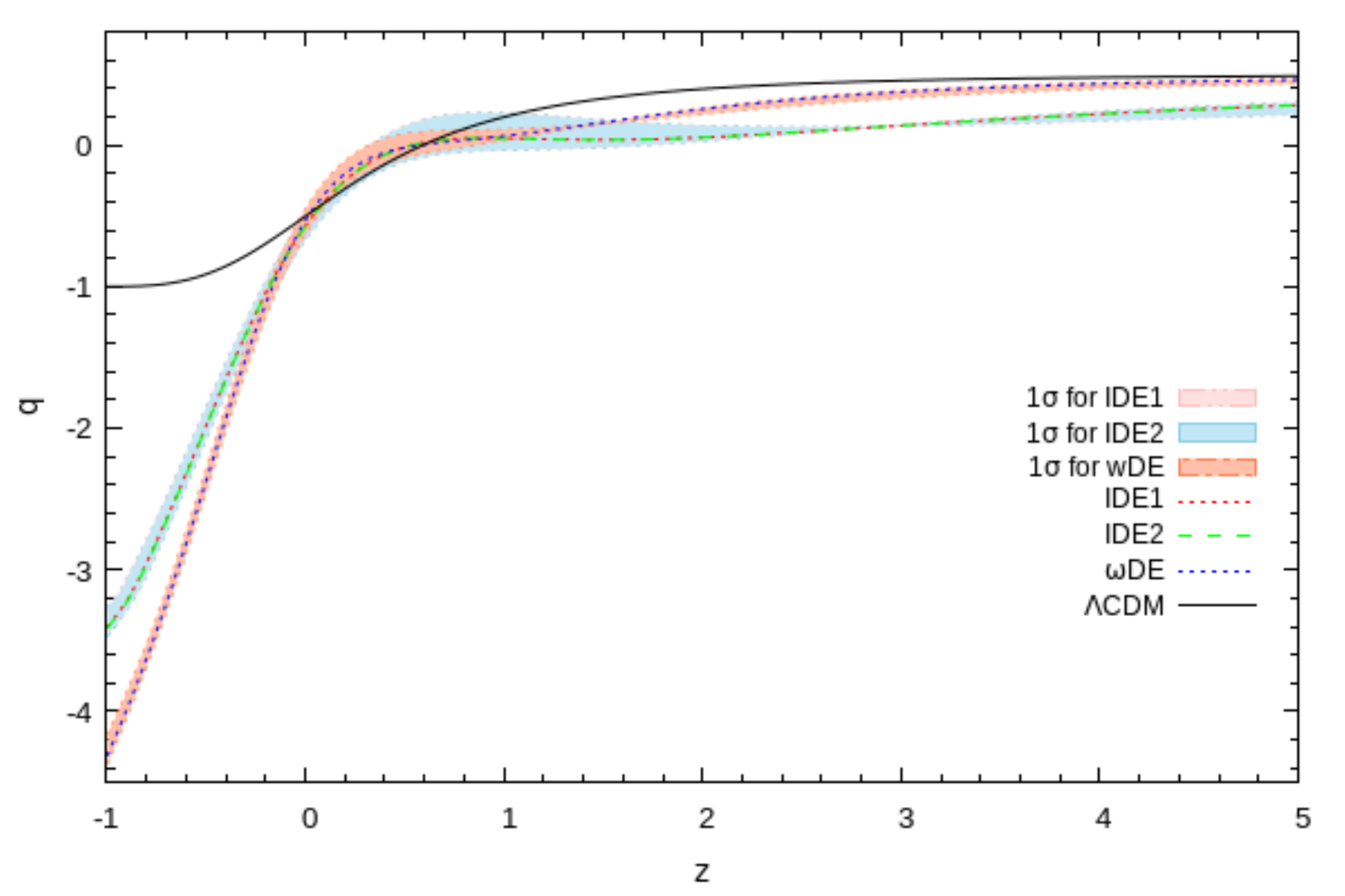}
 \end{tabular}} 
 \caption{\label{Reconstructions1} Shows the background evolution of $\mathrm{\omega_{DE}}$ (left above panel), ${\rm I}_{\rm Q}$ 
 (right above panel), $\mathrm{{\bf H}/H_{0}}$ (left below panel) and $\mathrm{\bf q}$ (right below panel) along $z$ for the present models. 
 Here, we have fixed the best-fit values of Table \ref{Bestfits} and have considered the respective constraints at $1\sigma$ error for the $\omega$DE and IDE models.} 
\end{figure*}
 \begin{figure*}[!htb]
 \centering 
 \resizebox{0.75\textwidth}{!}{
 \begin{tabular}{*{2}{ c }}
 \includegraphics{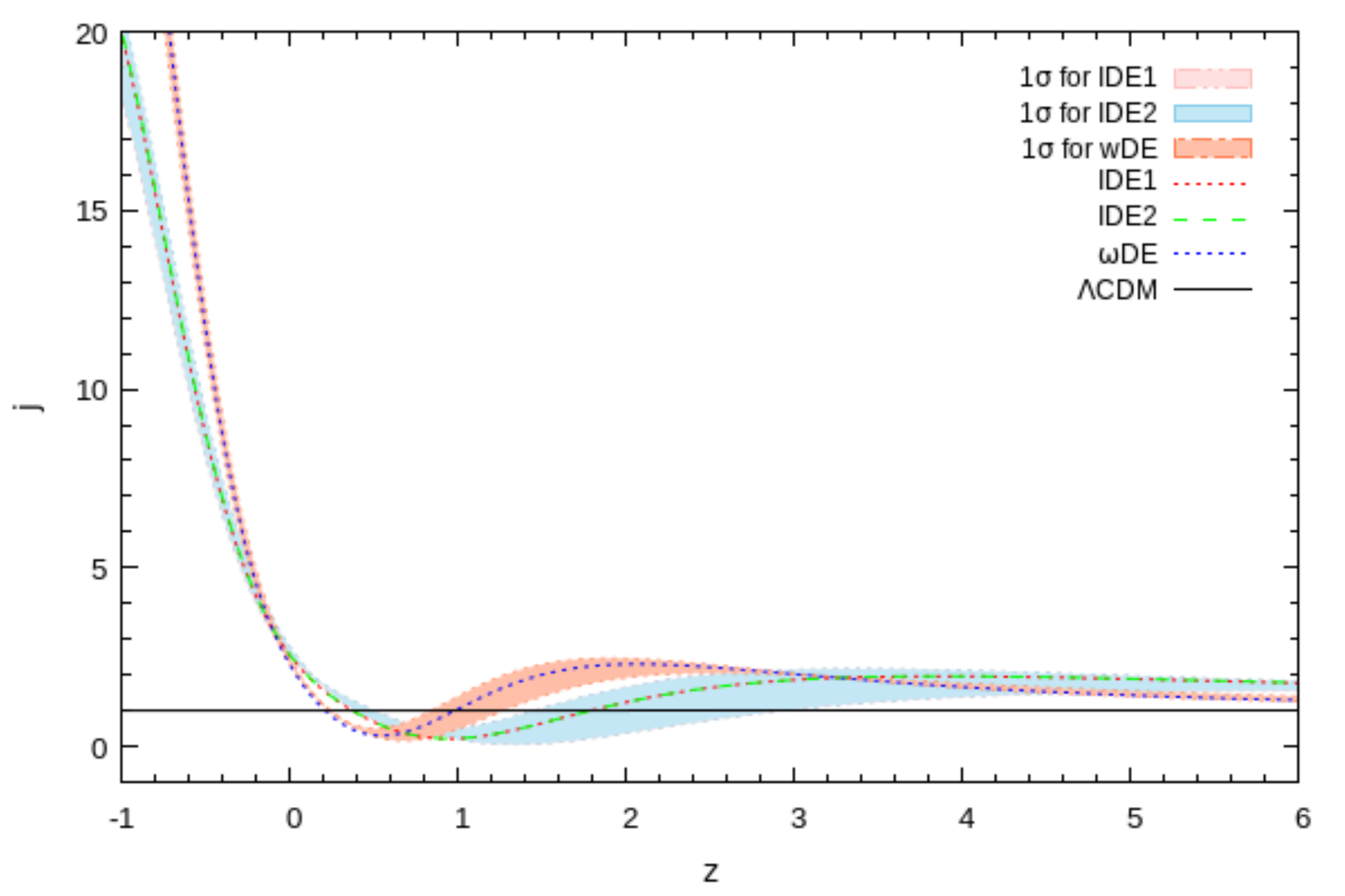}&\includegraphics{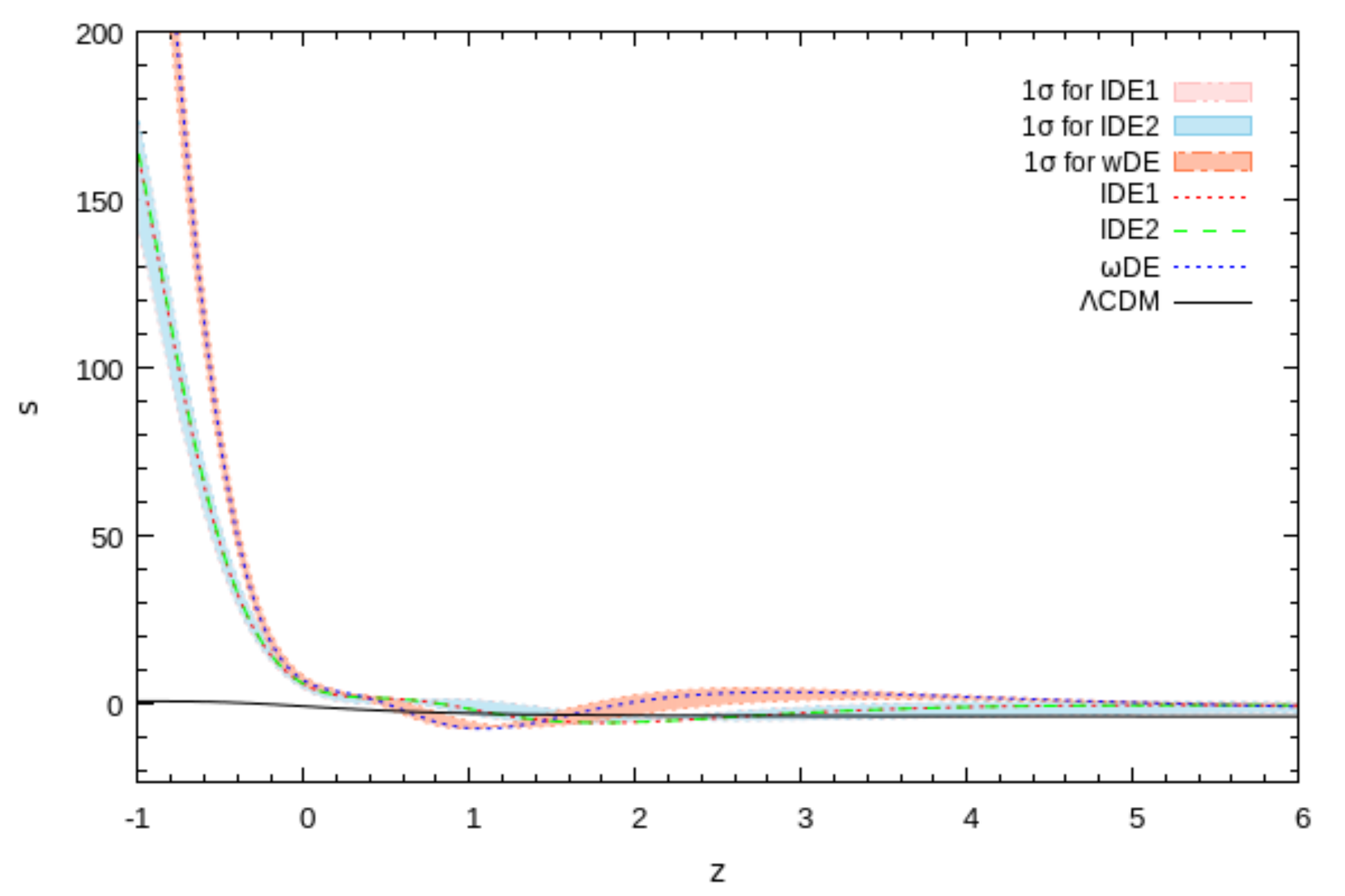}\\
 \includegraphics{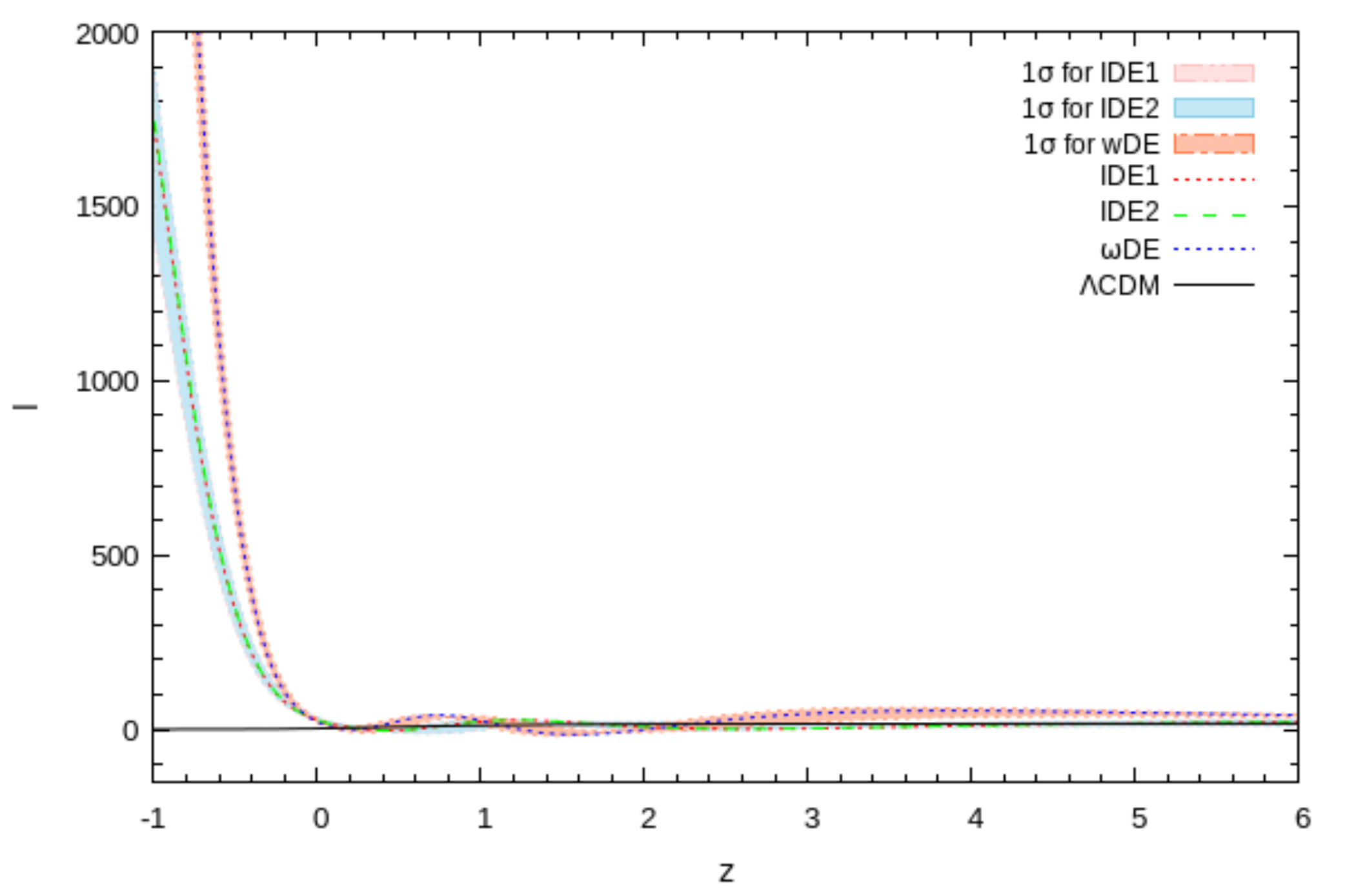}&\includegraphics{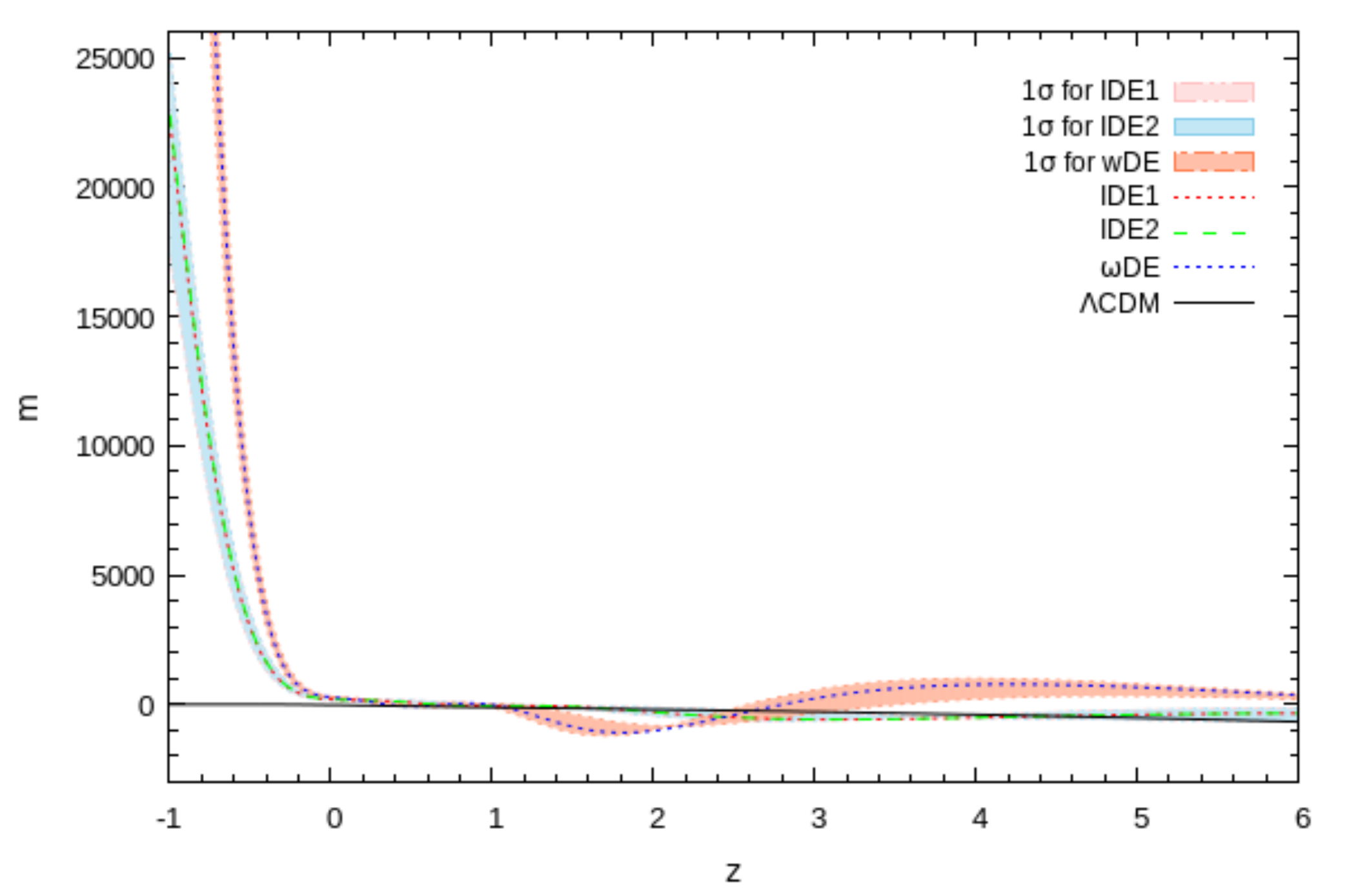}
 \end{tabular}} 
  \caption{\label{Reconstructions2} Displays the background evolution of $\mathrm{\bf j}$ (left above panel), $\mathrm{\bf s}$ (right above panel), 
 $\mathrm{\bf l}$ (left below panel) and $\mathrm{\bf m}$ (right below panel) along $z$ for the present scenarios. Here, we have fixed the best-fit 
 values of Table \ref{Bestfits} and have considered the respective constraints at $1\sigma$ error for the $\omega$DE and IDE models.}
 \end{figure*}
 \\
 The cosmic evolution of $\mathrm{\bf q}$, $\mathrm{\bf j}$, $\mathrm{\bf s}$ and $\mathrm{\bf l}$ along $z$ for the three scenarios 
 within $1\sigma$ confidence level and in the range $-1\leq z\leq 5$ are plotted in the left below panel of Fig. \ref{Reconstructions1} and in the panels of 
 Fig. \ref{Reconstructions2}. From the last panel of Figure \ref{Reconstructions1}, it is evident that $\mathrm{\bf q}$ shows a transition from decelerated phase to 
 accelerated phase at the transition redshift, $z_{t}$, defined from $\mathrm{\bf q}(z_{t})=0$. In Table \ref{Bestfitstoday}, we list the current best-fit values of the cosmographic parameters and the 
 best-fit values of $z_{t}$ (sixth row) within $1\sigma$ error for the three models. Among the three scenarios, the $\Lambda$CDM model presents a larger 
 $z_{t}$, and the IDE1 case presents a smaller $z_{t}$. Likewise, from this Table we can see that, for the three models, $z_{t}$ is located at $0.5<z_{t}<1.0$, which 
 is consistent at $1\sigma$ error with the results presented in \cite{Moresco2016,Gill2012,Ratra2013,Maurice2015,Parkinson2016,Xia2016,Madiyar2017,Motta2018}.\\  
 In addition, for the three models, we find that the $1\sigma$ confidence regions of $\mathrm{\bf q}$ are different in the future. While in the $\omega$DE model, 
 $\mathrm{\bf q}$ is slightly smaller at $z<0$ in comparison with the results found in the other models. In contrast, the amplitudes of 
 $\mathrm{\bf j}$, $\mathrm{\bf s}$, $\mathrm{\bf l}$ and  $\mathrm{\bf m}$ in the $\omega$DE model become larger and remain finite in the past or future when are 
 compared with those predicted in the other models. In the $\omega$DE and IDE models, these last parameters present an oscillatory behavior around the best fit curve 
 of the $\Lambda$CDM model and can change their sign at $z<3$. These effects may be a consequence of the chosen ansatzes for $Q$ and $\mathrm{\omega_{DE}}$.\\
 As we can see from Table \ref{Bestfitstoday}, the present values of the cosmographic parameters can be used to establish differences among the three DE models.
 \section{Conclusions}\label{SectionConclusions} 
In this work, we are interested in reconstructing the whole evolutionary histories of the first six cosmographic parameters from the past to future, 
allowing us to extract information about the kinematic state of the universe today. For this reason, we examined an interacting DE model (IDE) fills with 
two interacting components such as DM and DE, together with two non-interacting components decoupled from the dark sectors such as baryons and radiation. Here, 
we propose an interaction $Q$ proportional to the DM energy density, to the Hubble parameter $\mathrm{\bf H}$, and to a time-varying function, ${{\rm I}}_{\rm Q}$, 
expanded in terms of the Chebyshev polynomials $T_{n}$, defined in the interval $[-1,1]$. Besides, we also reconstruct a non-constant $\mathrm{\omega_{DE}}$, 
in function of those polynomials. These ansatzes have been proposed so that their cosmic evolution are free of divergences at the present and future times, 
respectively. Based on a combined analysis of geometric probes including JLA + BAO + CMB + H data and using the MCMC method, we constrain the parameter space 
and compared it with the results obtained from two different non-interacting models, presented in Tables \ref{Bestfits} and \ref{Bestfitstoday}, 
respectively.\\ 
Likewise, from Table \ref{Bestfits} and the upper panels of Fig. \ref{Reconstructions1}, our fitting results show that $\mathrm{\omega_{DE}}$ crosses $-1$ twice. 
Similarly, ${{\rm I}}_{\rm Q}$ can cross twice the line $Q=0$ as well. These crossing features are favored by the data at $1\sigma$ error. 
On the other hand, the combined impact of both $Q$ and $\mathrm{\omega_{DE}}$ on the evolution of the cosmographic parameters along $z$, 
are shown in the below panels of Fig. \ref{Reconstructions1} and in the panels of Fig. \ref{Reconstructions2}, respectively. From these panels, it 
has also found that the evolution of $\mathrm{{\bf H}/H_{0}}$, $\mathrm{\bf q}$, $\mathrm{\bf j}$, $\mathrm{\bf s}$, $\mathrm{\bf l}$ and $\mathrm{\bf m}$ in the IDE model deviates 
significantly from those inferred in the $\Lambda$CDM and $\omega$DE models, respectively, and moreover, they do not diverge in a far future, except for 
the behavior of $\mathrm{{\bf H}/H_{0}}$. It meant that, these detected deviations are brought about mainly by $Q$ and $\mathrm{\omega_{DE}}$. Thus, these 
reconstructed cosmographic parameters are sensitive to the evolution of $Q$ and $\mathrm{\omega_{DE}}$, respectively.
Furthermore, the right below panel of Fig. \ref{Reconstructions1} indicates that in the $\omega$DE and IDE models, the universe is less accelerated in a far 
future respect to the predicted value by the $\Lambda$CDM model.
According to the results presented previously, the $\mathrm{\bf j}$, $\mathrm{\bf s}$, $\mathrm{\bf l}$ and $\mathrm{\bf m}$ parameters found in the IDE model exhibit 
qualitatively different behaviors when are compared with those obtained in the $\Lambda$CDM or $\omega$DE models. 
For instance, these effects can be understood by considering the extra-terms $\mathrm{\lambda_{0}}$, $\mathrm{\lambda_{1}}$ and $\mathrm{\lambda_{2}}$ in the DM energy density, 
(see Eq. (\ref{hubble})), which increases and, in consequence, amplifies the amount of DM. Similarly, $\mathrm{\omega_{0}}$, $\mathrm{\omega_{1}}$ 
and $\mathrm{\omega_{2}}$ affect $\mathrm{\omega_{DE}}$. As a result, $\mathrm{R}$, and in consequence, the respective $\mathrm {\bf q}$ in the $\omega$DE and IDE 
models become lesser than that inferred in the $\Lambda$CDM model.\\
In addition, we also see that our numerical estimations agree within $1\sigma$ error with those obtained in 
\cite{Xu-Li-Lu2009,Wang-Dai-Qi2009,Vitagliano-Viel2010,Xu2011,Gruber2012,Sendra2013,Gruber2014,Movahed2017,Pan2018,Ming2017,Rodrigues2018,Ruchika2018}. We have confirmed that the ansatzes for $Q$ and $\mathrm{\omega_{DE}}$ in terms of $z$ are successful and valid 
to reconstruct the cosmographic series. In this sense, the IDE model can be compared and distinguished from the $\Lambda$CDM and $\omega$DE models, by using the present 
values of the cosmographic parameters, given by Table \ref{Bestfitstoday}.
We believe that the two ansatzes proposed for $Q$ and $\mathrm{\omega_{DE}}$ in terms of the Chebyshev polynomials are very successful to explore 
the dynamical evolution of DE and have shown that they can be employed to reconstruct the first six cosmographic parameters. We suggest that those 
ansatzes should be further investigated.\\\\
\begin{scriptsize}
\textbf{\textsf {Acknowledgments}} both
\textsf{The author is indebted to the Institute of Physics and Mathematics (IFM-UMSNH) for its hospitality and support.}
\end{scriptsize}
   

\begin{thebibliography}{72}
\begin{scriptsize}
\bibitem{Conley2011}
Conley A et al., \textit{Astrophys. J. Suppl.} \textbf{192} (2011) 1.
\bibitem{Jonsson2010}
J\"onsson, J., et al., \textit{Mon. Not. Roy. Astron. Soc.} \textbf{405} (2010) 535.
\bibitem{Betoule2014}
Betoule M et al., \textit{Astron. and Astrophys.} \textbf{568} (2014) A22.
\bibitem{Planck2015}
Planck 2015 results, XIII. Cosmological parameters, \textit{Astron. Astrophys.} \textbf{594} (2016) A13.
\bibitem{Hinshaw2013} 
WMAP collaboration, G. Hinshaw et al., \textit{Astrophys. J. Suppl.} \textbf{208} (2013) 19.
\bibitem{Beutler2011}
F. Beutler et al., \textit{Mon. Not. Roy. Astron. Soc.} \textbf{416} (2011) 3017.
\bibitem{Ross2015}
A. J. Ross et al., \textit{Mon. Not. Roy. Astron. Soc.} \textbf{449} (2015) 835.
\bibitem{Percival2010}
W. J. Percival et al., \textit{Mon. Not. Roy. Astron. Soc.} \textbf{401} (2010) 2148.
\bibitem{Blake2011}
Blake C. et al., \textit{Mon. Not. Roy. Astron. Soc.} \textbf{415} (2011) 2876; \textit{Mon. Not. Roy. Astron. Soc.} \textbf{418} (2011) 1725.
\bibitem{Kazin2010}
E. A. Kazin et al., \textit{Astrophys. J.} \textbf{710} (2010) 1444.
\bibitem{Anderson2014a}
L. Anderson et al., \textit{Mon. Not. Roy. Astron. Soc.} \textbf{441} (2014) 24.
\bibitem{Padmanabhan2012}
N. Padmanabhan et al., \textit{Mon. Not. Roy. Astron. Soc.} \textbf{427} (2012) 2132.
\bibitem{Chuang2013a}
C. H. Chuang and Y. Wang, \textit{Mon. Not. Roy. Astron. Soc.} \textbf{435} (2013) 255.
\bibitem{Chuang2013b}
C-H. Chuang and Y. Wang, \textit{Mon. Not. Roy. Astron. Soc.} \textbf{433} (2013) 3559.
\bibitem{Debulac2015}  
T. Delubac et al., \textit{Astron. Astrophys.} \textbf{574} (2015) A59.
\bibitem{FontRibera2014} 
A. Font-Ribera et al., \textit{J. Cosmol. Astropart. Phys.} \textbf{05} (2014) 027.
\bibitem{Eisenstein1998}
D. J. Eisenstein, W. Hu, \textit{Astrophys. J.} \textbf{496} (1998) 605.
\bibitem{Eisenstein2005}
D. J. Eisenstein et al., \textit{Astrophys. J.} \textbf{633} (2005) 560.
\bibitem{Hemantha2014}
M. D. P. Hemantha, Y. Wang and C-H. Chuang., \textit{Mon. Not. Roy. Astron. Soc.} \textbf{445} (2014) 3737.
\bibitem{Bond-Tegmark1997}
J. R. Bond, G. Efstathiou and M. Tegmark, \textit{Mon. Not. Roy. Astron. Soc.} \textbf{291} (1997) L33.
\bibitem{Hu-Sugiyama1996}
W. Hu and N. Sugiyama, \textit{Astrophys. J.} \textbf{471} (1996) 542.
\bibitem{Neveu2016}
J. Neveu, V. Ruhlmann-Kleider, P. Astier, M. Besan\c con, J. Guy, A. M\"oller, E. Babichev, \textit{Astron. and Astrophys.} \textbf{600} (2017) A40.
\bibitem{Zhang2014} 
C. Zhang et al., \textit{Res. Astron. Astrophys.} \textbf{14} (2014) 1221.
\bibitem{Simon2005}
J. Simon, L. Verde and R. Jimenez, \textit{Phys. Rev.} \textbf{D 71} (2005) 123001.
\bibitem{Moresco2012}
M. Moresco et al., \textit{J. Cosmol. Astropart. Phys.} \textbf{8} (2012) 006.
\bibitem{Moresco2016}
M. Moresco, L. Pozzetti, et al., \textit{J. Cosmol. Astropart. Phys.} \textbf{05} (2016) 014.
\bibitem{Gastanaga2009}
E. Gasta\~naga, A. Cabre, L. Hui, \textit{Mon. Not. Roy. Astron. Soc.} \textbf{399} (2009) 1663.
\bibitem{Oka2014}
A. Oka et al., \textit{Mon. Not. Roy. Astron. Soc.} \textbf{439} (2014) 2515.
\bibitem{Blake2012}
C. Blake et al., \textit{Mon. Not. Roy. Astron. Soc.} \textbf{425} (2012) 405.
\bibitem{Stern2010}
D. Stern, R. Jimenez, L. Verde, M. Kamionkowski and S. A. Stanford, \textit{J. Cosmol. Astropart. Phys.} \textbf{02} (2010) 008.
\bibitem{Moresco2015}
M. Moresco, \textit{Mon. Not. Roy. Astron. Soc.} \textbf{450} (2015) L16-L20.
\bibitem{Busca2013}
N. G. Busca et al., \textit{Astron. Astrophys.} \textbf{552} (2013) A96.
\bibitem{DES2006}
V. Sahni, \textit{Lect. Notes Phys.} \textbf{653} (2004) 141; E. J. Copeland, M. Sami and S. Tsujikawa, \textit{Int. J. Mod. Phys.} \textbf{D 15} (2006) 1753.
\bibitem{OptionsDE}
U. Seljak, et al., \textit{Phys. Rev.} \textbf{D 71} (2005) 103515;
M. R. Garousi, M. Sami, and S. Tsujikawa, \textit{Phys. Rev.} \textbf{D 71} (2005) 083005.
M. K. Mak and T. Harko, \textit{Phys. Rev.} \textbf{D 71} (2005) 104022;
X. Cheng, Y. Gong and E. N. Saridakis, \textit{J. Cosmol. Astropart. Phys.} \textbf{04} (2009) 001;
E. Rozo et al., \textit{Astrophys. J.} \textbf{708} (2010) 645.
\bibitem{modifiedDE}
G. R. Dvali, G. Gabadadze, M. Porrati, \textit{Phys. Lett.}\textbf{B 485} (2000) 208;
C. Deffayet, G. R. Dvali, G. Gabadadze, \textit{Phys. Rev.} \textbf{D 65} (2002) 044023;
S. M. Carroll, V. Duvvuri, M. Trodden, M. S. Turner, \textit{Phys. Rev.} \textbf{D 70} (2004) 043528;
M. Li, X.-D. Li, S. Wang and Y. Wang, \textit{Commun. Theor. Phys.} \textbf{56} (2011) 525;
Poplawski, N. J, \textit{Phys. Lett.} \textbf{B 640} (2006) 135;
S. Capozziello, E. Elizalde, E. Noriji, S. D. Odintsov \textit{Phys. Lett.} \textbf{B 671}, (2009) 193;
L. Xu, W. Li and J. Lu. Shen, \textit{Mod. Phys. Lett.} \textbf{A 24} (2009) 1355;
X. Zhang,\textit{Phys. Rev. D} \textbf{79} (2009) 103509;
C. J. Feng and X. Z. Li, \textit{Phys. Lett.} \textbf{B 680} (2009) 184;
M. Li, X. D. Li, S. Wang and X. Zhang,\textit{J. Cosmol. Astropart. Phys.} \textbf{06} (2009) 036;
L. Xu, J. Lu. Shen, and  W. Li, \textit{Eur. Phys. J.} \textbf{C. 64} (2009) 89;
C. J. Feng and X. Z. Li, \textit{Phys. Lett.} \textbf{B 680} (2009) 355, \textbf{679} (2009) 151;
S. B. Cheng and J. L. Jing, \textit{arxiv: 0904.2950};
C. J. Feng and X. Zhang, \textit{Phys. Lett.} \textbf{B 680}(2009) 399. 
\bibitem{Weinberg1972}  
S. Weinberg, \textit{Cosmology and gravitation. John Wiley Sons. New York. USA. 1972}
\bibitem{Visser2005}
M. Visser, \textit{Gen. Rel. Grav.} \textbf{37} (2005) 1541.
\bibitem{Xu-Li-Lu2009}
 L. Xu, W. Li and J. Lu, \textit{J. Cosmol. Astropart. Phys.} \textbf{07} (2009) 031.
\bibitem{Wang-Dai-Qi2009}
F. Y. Wang, Z. G. Dai, and Shi Qi, \textit{Astron. Astrophys.} \textbf{507} (2009) 53-59.
\bibitem{Vitagliano-Viel2010}
 V. Vitagliano, J. Q. Xia, S. Liberati and M. Viel, \textit{J. Cosmol. Astropart. Phys.} \textbf{03} (2010) 005. 
\bibitem{Xu2011}
L. Xu, and Y. Wang, \textit{Phys. Lett.} \textbf{B 702} (2011) 114-120.
\bibitem{Gruber2012}
A. Aviles, C. Gruber, O. Luongo, H. Quevedo, \textit{Phys. Rev.} \textbf{D 86} (2012) 123516.
\bibitem{Sendra2013}
R. Lazkoz, J. Alcaniz, C. Escamilla-Rivera, V. Salzano, I. Sendra, \textit{J. Cosmol. Astropart. Phys.} \textbf{12} (2013) 005.
\bibitem{Gruber2014}
C. Gruber and O. Luongo, \textit{Phys. Rev.} \textbf{D 89} (2014) 103506.
\bibitem{Movahed2017}
B. Mostaghel, H. Moshafi, and S.M.S. Movahed, \textit{Eur. Phys. J.} \textbf{C. 77} (2017) 541.
\bibitem{Pan2018}
S. Pan, A. Mukherjee, and N. Banerjee., \textit{Mon. Not. Roy. Astron. Soc.} \textbf{477} (2018) 1189.
\bibitem{Ming2017}
Ming-Jian Zhang,Hong Li and Jun-Qing Xia, \textit{Eur. Phys. J.} \textbf{C 77} (2017) 434.
\bibitem{Rodrigues2018}
C. Rodrigues Filho, Edesio M. Barboza Jr., \textit{J. Cosmol. Astropart. Phys.} \textbf{07} (2018) 037.
\bibitem{Ruchika2018}
Salvatore Capozziello, Ruchika, and Anjan A. Sen, \textbf{arxiv: 1806.03943 v2}.
\bibitem{Interacting}
Z. K. Guo, N. Ohta, and S. Tsujikawa, \textit{Phys. Rev.} \textbf{D 76} (2007) 023508; 
J. H. He and B. Wang, \textit{J. Cosmol. Astropart. Phys.} \textbf{06} (2008) 010;
S. Campo, R. Herrera and D. Pav\'on, \textit{J. Cosmol. Astropart. Phys.} \textbf{01} (2009) 020;
S. Cao, N. Liang and Z. H. Zhu, \textit{Int. J. Mod. Phys.} \textbf{D 22} (2013) 1350082.
\bibitem{Pavons}
D. Pav\'on, B. Wang, \textit{Gen.Rel.Grav.} \textbf{41} (2009) 1-5;
S. del Campo, R. Herrera, G. Olivares, and D. Pav\'on, \textit{Phys. Rev.} \textbf{D 74} (2006) 023501.
\bibitem{Wangs}
B. Wang, C. Y. Lin and E. Abdalla, \textit{Phys. Lett.} \textbf{B 637} (2006) 357.
\bibitem{Cueva-Nucamendi2012}
F. Cueva Solano and U. Nucamendi, \textit{J. Cosmol. Astropart. Phys.} \textbf{04} (2012) 011;
F. Cueva Solano and U. Nucamendi, \textbf{arXiv: 1207.0250}.
\bibitem{valiviita2008}
J. Valiviita, E. Majerotto and R. Maartens, \textit{J. Cosmol. Astropart. Phys.} \textbf{07} (2008) 020.
\bibitem{Clemson2012}
T. Clemson, K. Koyama, G. B. Zhao, R. Maartens and J. Valiviita \textit{Phys. Rev.} \textbf{D 85} (2012) 043007.
\bibitem{Chevallier-Linder}
M. Chevallier, D. Polarski, \textit{Int. J. Mod. Phys.} \textbf{D 10} (2001) 213;
E. V. Linder, \textit{Phys. Rev. Lett.} \textbf{90} (2003) 091301.  
\bibitem{Li-Ma}
H. Li and X. Zhang, \textit{Phys. Lett.} \textbf{B 703} (2011) 119;
J. Z. Ma and X. Zhang, \textit{Phys. Lett.} \textbf{B 699} (2011) 233. 
\bibitem{Martinez2008}{Crossings}
E. F. Martinez and L. Verde, \textit{J. Cosmol. Astropart. Phys.} \textbf{08} (2008) 023.
\bibitem{Campo-Herrera2015}
S. del Campo, R. Herrera, and D. Pav\'on \textit{Phys. Rev.} \textbf{D 91} (2015) 123539.
\bibitem{Ratios}
L. P. Chimento, A. S. Jakubi, D. Pav\'on, and W. Zimdahl, \textit{Phys. Rev.} \textbf{D 67} (2003) 083513;
J. Q. Xia and M. Viel, \textit{J. Cosmol. Astropart. Phys.} \textbf{04} (2009) 002.
\bibitem{Lewis2002}
A. Lewis and S. Bridle, \textit{Phys. Rev.} \textbf{D 66} (2002) 103511;\\ 
\textit{http://cosmologist.info/cosmomc/}.
\bibitem{Sharov2015}
G. S. Sharov, \textit{J. Cosmol. Astropart. Phys.} \textbf{06} (2016) 023.
\bibitem{Nesseris2007}
S. Nesseris and L. Perivolaropoulos, \textit{J. Cosmol. Astropart. Phys.} \textbf{01} (2007) 018.
\bibitem{Cai2010}
R. G. Cai and Q. Su, \textit{Phys. Rev.} \textbf{D 81} (2010) 103514.
\bibitem{Li2011}
Y. H. Li and X. Zhang, \textit{Eur. Phys. J.} \textbf{C 71} (2011) 1700.
\bibitem{Gill2012}
J. A. S. Lima, J. F. Jesus, R. C. Santos, and M. S. S. Gill, \textbf{arXiv: 1205.4688}.
\bibitem{Ratra2013}
Omer Farooq, Sara Crandall, Bharat Ratra, \textit{Phys. Lett.} \textbf{B 726} (2013) 72-82.
\bibitem{Maurice2015}
Maurice H. P. M. van Putten, \textit{Mon. Not. Roy. Astron. Soc.} \textbf{450} (2015) L48.
\bibitem{Parkinson2016}
D. Muthukrishna and D. Parkinson, \textit{J. Cosmol. Astropart. Phys.} \textbf{11} (2016) 052.  
\bibitem{Xia2016}
Ming-Jian Zhang and Jun-Qing Xia, \textit{J. Cosmol. Astropart. Phys.} \textbf{12} (2016) 005.
\bibitem{Madiyar2017}
O. Farooq, F. Madiyar, S. Crandall, B. Ratra, \textit{Astrophys. J.} \textbf{835} (2017) 01.
\bibitem{Motta2018}
J. Rom\'an-Garza, T. Verdugo, J. Maga\~na, V. Motta, \textbf{arXiv: 1806.03538v1}.
\end{scriptsize}
\end{thebibliography}
\end{document}